\documentclass[12pt]{article}

\usepackage{amssymb,graphicx}

\newcommand{\be}{\begin{equation}}
\newcommand{\ee}{\end{equation}}
\newcommand{\bea}{\begin{eqnarray}}
\newcommand{\eea}{\end{eqnarray}}

\begin{document}
\begin{titlepage}

\begin{flushright}
{\today}
\end{flushright}
\vspace{1in}

\begin{center}
\Large
{\bf  Analyticity Properties  of Scattering
Amplitude  in Theories with Compactified Space Dimensions: The Proof
of Dispersion Relations}\footnote{ Dedicated in fond memories  of  
 Andr\'e Martin.       }
\end{center}

\vspace{.2in}

\normalsize

\begin{center}
{ Jnanadeva Maharana   \\
E-mail maharana$@$iopb.res.in}\\ 
\end{center}

\normalsize

\begin{center}
 {\em Institute of Physics \\
NISER\\
Bhubaneswar - 751005, India \\ 
Invited Review Article for International Journal of Modern Physics A
     }

\end{center}

\vspace{.2in}

\baselineskip=24pt

\begin{abstract}
The analyticity properties of the scattering amplitude for a  
massive scalar field is reviewed in this article
where the spacetime geometry is $R^{3,1}\otimes S^1$ i.e.  one spatial 
dimension is compact.
 Khuri investigated  
the analyticity of scattering amplitude
in a nonrelativitstic potential model in three dimensions with an additional
 compact dimension. He showed that, under
certain circumstances, the forward amplitude is nonanalytic. He argued that 
in high
energy scattering  if such a behaviour persists it would be 
in conflicts with the established results of quantum field theory and LHC
might observe such behaviors.
We envisage a real scalar massive field in flat Minkowski spacetime
in five dimensions. The Kaluza-Klein (KK) compactification is implemented
on a circle. The resulting four dimensional manifold is
$R^{3,1}\otimes S^1$. The LSZ  formalism is adopted to
study the analyticity of the scattering amplitude. The
nonforward dispersion relation is proved. In addition the Jin-Martin bound
and an analog of the Froissart-Martin bound are proved. A novel proposal 
is presented to look for evidence of the large-radius-compactification
scenario.  A seemingly violation
of Froissart-Martin bound at LHC energy might hint that
an extra  dimension might be decompactified. However, we find no evidence
for violation of the bound in  our analysis.

\end{abstract}

\vspace{.5in}

\end{titlepage}


\section{Introduction }

The purpose of this review is to present the study of the 
analyticity properties
of scattering amplitude for  a massive hermitian scalar field theory
in four dimensional spacetime with one additional compactified
spatial dimension. This extra dimension is a circle i.e. we consider
$S^1$ compactification. The axiomatic field theory formulation of
Lehmann, Symanzik and Zimmermann (LSZ) is adopted to prove the
dispersion relations for the four point amplitude. Our motivation
to undertake this investigation stems from several discussion
with Andr\'e Martin in 2018. Khuri, in 1995, had studied the
analyticity property of forward scattering amplitude in a
nonrelativistic potential model in three dimensions with an extra
compact spatial dimension (he considered $S^1$ compactification). 
 He concluded
that, under certain circumstances, as will be elaborated later, the
amplitude does not satisfy the same  analyticity property  
as enjoyed by the amplitudes
of conventional potential models in three dimensional space. Martin
raised an important question: What are the analyticity properties of
an amplitude in
a relativistic quantum field theory where one spatial dimension is compact?
Should the established analyticity properties of an 
amplitude, derived rigorously  in
 field theories in $D=4$,  be violated in the
compactified-spatial-dimension theory it would lead to grave consequences.
The fallout would be that some of the fundamental axioms of local
quantum field theories would be questioned. We had undertaken  an
investigation to address these issues. The details will be elaborated in
the sequel. This article is presented in a pedagogical style. Thus
the reader, with a background in relativistic quantum field theory,
can work out the steps if interested. The approach presented here
for the problem at hand, to the
best of the knowledge of the author, has not been reported previously.
Therefore, we have expended on our earlier investigations \cite{jm1,jm2}
in order
to make this article accessible to a wider audience. We begin with
a few remarks to motivate the reader.
\\
There are two approached to study scattering processes. In the perturbative
formulation, we decompose the Lagrangian into a free part and an 
interaction part. The free field equations are exactly solvable. The
procedures of perturbation theory allow us to compute the S-matrix
elements order by order. We encounter divergences in computations.
The renormalization prescription consistently removes the infinities at
each order.  Therefore, in this approach, the renormalizable theories are
able to give us finite results which are subjected to
verifications against experimental data.  The crossing symmetry is maintained
at each order as we include all Feynman diagrams.  In other words,
when we consider all the Feynman diagrams in a  given order, we
include all direct channel diagrams and all possible crossed channel diagrams.
The diagrammatic technique already has built in crossing symmetry.
Furthermore, the
unitarity property of the amplitude is to be ensured order by order in the
perturbation expansions. There are well laid down prescriptions to
test the analyticity properties. The success of quantum electrodynamic (QED)
in computing the anomalous magnetic moments of
charged leptons and in computing Lamb shift with unprecedented accuracy tell
the success of perturbative formulation of renormalizable
quantum field theories. Moreover, the predictions of the
i standard model of particle physics
has been tested to great degree of accuracy against experimental data.
\\
The S-matrix proposal of Heisenberg \cite{heisenberg} is 
radically different  from the perturbation
theoretic formulations. He argued that in a scattering experiment the
initial states are in the remote past. The projectile hits the target and
the experimentalists observe the out going particles in remote
future. Therefore, the initial state consists of free particles characterized
by their physical mass, angular momenta and spins etc. Similarly, the
attributes of final
states are observed in the detector. We may imagine the initial state to
be a vector which is prepared and final state to be another vector. However,
each complete set of vectors span the Hilbert space. Therefore, there must
be a unitary operator connecting the two sets in order to ensure the
conservation of probability. He designated it as the scattering matrix or
S-matrix. The initial and final states contain particles whose masses
are observed quantities. Therefore, there are no divergence difficulties as
encountered in perturbative approach. This is a very naive and qualitative
way of introducing the concept of his S-matrix. Thus there is a different
philosophical approach. The idea of Heisenberg was built on rigorous foundation
in subsequent years. The axiomatic formulation of  Lehmann, Symanzik and
Zimmermann (LSZ) \cite{lsz}
 is a landmark in relativistic quantum field theories (QFT).
Wightman \cite{wight}  proposed that field theories be studied
in terms of vacuum
expectation values of product of field operators and he introduced a set of
axioms. There are very important theorems
on the analyticity properties of scattering amplitude
which have been proved from the frameworks of
general axiomatic  field theories 
\cite{book1,book2,book3, fr1,lehm1,sommer,eden,wight,jost,streat,kl,ss,bogo}.
\\                
We shall adopt the axiomatic approach of LSZ in this article and
this formulation will be elaborated in the next section.
They introduced the notion of asymptotic fields and interacting fields.
Moreover, a Lagrangian is not introduced explicitly. Therefore, there is
no proposal of separating the theory into a free and an interacting theory.
Some of the axioms include existence of a Hilbert space, Lorentz invariance,
microcausality to mention a few (see next section for details). 
It is quite startling that the formulation
enables computation of a scattering amplitude. Furthermore, the axioms
lead to a set of linear relationships among various amplitudes. It is the
hallmark of the axiomatic field theory. It must be emphasized  that,
within this linear framework,  several important attributes of the
scattering amplitude are derived. Of special importance is the
proof of dispersion relations. As we shall discuss in subsequent sections,
the dispersion relations are derived from general requirements such as
Lorentz invariance and microcausality. Indeed, there is a deep and intimate
relationship between analyticity of scattering amplitude and causality
when we consider local quantum field theories. The unitarity of the S-matrix,
which
is a nonlinear relation, is not invoked in order to arrive at the linear
relations. We mention {\it en pasant} that the unitarity of S-matrix is
proved in the LSZ framework (\cite{ss,nishijima}). Indeed, the
analyticity properties of scattering amplitudes are derived rigorously in
the LSZ formalism. One of most celebrated result is the Froissart-Martin
bound \cite{fr,andre} on total cross sections
\bea
\label{f-m}
\sigma_t(s)\le {{4\pi}\over{ t_0}}log^2({s\over{s_0}})
\eea
The essential interpretation of the bound is as follows. We may ask what is
the energy dependence of a total cross section, $\sigma_t(s)$, at asymptotic
energies? The above bound implies that  its energy dependence is not
arbitrary. Moreover, the constant prefactor appearing in the $r.h.s.$ 
of (\ref{f-m}) was determined from first
principles by Martin \cite{andre}. We need to introduce 
a constant, $s_0$, in order
that the argument of the $log$ is dimensionless. However, it cannot be
 determined from first principles. The necessary ingredients for derivation
of the upper bound are: (i) Analyticity of the scattering amplitude, (ii)
polynomial boundedness of the scattering amplitude as the energy
tends to asymptotic values and (iii) the unitarity  bounds on partial wave
amplitudes. All the three properties have been proved from axiomatic
field theories. The total cross sections, $\sigma_t$, measured in high
energy experiment
respect the Froissart-Martin bound.
Should there be  conclusive experimental
evidence
of the violation of this bound the fundamental axioms of local relativistic
quantum field theories
would be questioned. There is a  host of results,
usually presented as upper and
lower bounds on
experimentally measurable parameters, which have been subjected to experimental tests.
There are no evidences for violation of any these rigorous bounds
\cite{book2,eden,roy}.\\
We live in four spacetime dimensions. All the experiments are carried out in
laboratories and the
theories are constructed in four dimensions. The fundamental theories have
been subjected to
experimental tests. It is now an accepted idea that there might exist deeper
 fundamental
theories which live
in higher spacetime dimensions, $D>4$. There are well defined
theories such as
supersymmetric theories, supergravity theories and string theories which are
defined in higher
dimensions. The string theories hold the prospect of unifying the four
fundamental interactions.
Considerable attentions have been focused on string theories over last
few decades. Therefore,
it is pertinent to ask what relevance these theories have for physics in
four spacetime dimensions.
The proposal of Kaluza and Klein (KK) \cite{k1,k2} 
are invoked in the context of
such higher dimensional
theories. Kaluza and Klein \cite{k1,k2} envisaged a
  five dimensional theory of pure gravity
which is a generalized version
of Einstein's theory. They argued that the $5^{th}$ dimension is
compactified on a circle. Therefore,
the length scale probed in that era cannot resolve the size of the
extra compact dimension. They carried out
what is now known as the KK compactification scheme. It was shown that
the effective four dimensional
theory, in its massless sector, corresponds to a Maxwell-Einstein theory
in four dimensions. Moreover, there
is a tower of massive states (the KK states) and the mass in each level is
proportional to $1\over R$, where
$R$ is the radius of the circle. If the radius of the circle is very small
 then these  states become very massive and they cannot be
observed by the experimental techniques prevailing those days. 
The proposal of Kaluza and
Klein were employed to the
compactification of supergravity theories in 1970's (\cite {ss1}).  There was
a lot of interest in the KK compactification
after the second superstring revolution. In a rapid development, more
elegant and sophisticated compactification
schemes were developed \cite{gsw,polchi}. 
In the early phase of the string compactification era,
it was generally believed that the
radius of compactification of the compact dimensions would be in the vicinity
of the Planck length.
Therefore, the string excitations of the compactified theory
would be so heavy that their observation
will be out of reach of any
accelerator. Antoniadis \cite{anto93} proposed a scenario
where the compactification
scale is in the TeV range and
therefore, the KK excitations associated with string theories might
be detected in future accelerators. Antoniadis,
Munoz and Quiros \cite{amq} pursued this idea further.
Arkani-Hamed, Dimopoulos and Dvali \cite{add} proposed a large-radius-compactification
scheme of theirs and worked out
the phenomenological implications. Subsequently, Antoniadis, Arkani-Hamed,
Dimopoulos and Dvali \cite{aadd} advanced the idea
of large radius compactification proposal further. There were a lot of
activities, in subsequent years, to investigate
details of phenomenology of these proposals. The LHC was going  to be
commissioned in near future.
There was optimism that KK states of string would be observed at LHC and
it would be an experimental
confirmation of the ideas of string theories. A review of the
theoretical progress in this direction will be found in \cite{anto,luest}.
So far the
LHC experiments have established only limits on the scale
of compactification in the light of the large radius compactification
paradigm \cite{tev1,tev2}.\\
In an interesting paper Khuri \cite{khuri1},
investigated the analyticity of scattering amplitude
where a spatial dimension is compactified on a circle. He envisaged a
nonrelativistic potential model in three
spatial dimensions and with an extra compact spatial dimension.
The perturbative Greens function technique was
employed to compute the quantum mechanical scattering amplitude.
The additional feature was the existence of KK states. Thus the
standard integral equations which are employed in potential models of
scattering were modified appropriately.  In the
conventional study of scattering,  we consider incoming plain waves before
the scattering. The  Green's function formalism
enables us to extract the scattering amplitude. For the problem at hand,
the wave function is characterized by its
momentum, $\bf k$ and an integer,
$n$ due to the presence of a compact coordinate, $\Phi$;
$n$ is interpreted as the KK quantum number. Therefore, the initial
state is designated as $({\bf k}~ and ~n)$  evolving to a final state
$({\bf k}'~ and ~n')$. Moreover, there are conservation
laws which are to be respected. Khuri found that when one considers
the scattering process where  a state $({\bf k}, n=0)$
scatters into $({\bf k}', n'=0)$ then the scattering amplitude exhibits
the analyticity properties which are known for
a long time \cite{khuri2,gw,wong}. The situation is
different when one considers the process
$({\bf k}, n)\rightarrow ({\bf k}', n')$. The Green's function technique
developed by Khuri was employed. He showed that
for the scattering process $({\bf k}, n=1)\rightarrow ({\bf k}', n=1)$,
 the forward scattering amplitude exhibits a nonanalytic behavior
when it is computed to second order. In other words, when one considers
scattering states having momentum $|{\bf k}|$ and KK
quantum number $n=1$, the forward amplitude develops nonananytic
behavior whereas the amplitude for scattering in the $n=0$
sector exhibits no such attribute. Moreover, Khuri \cite{khuri1} remarked that
this phenomena will have very serious consequences if such
KK states are produced in the LHC experiments. Indeed, he cited the works
on Antoniadis \cite{anto93} and argued that the KK states
might be produced in the large-compactification-radius scenario.
It must be emphasized that the rigorous results of
Khuri was derived in the frameworks of nonrelativistic quantum mechanics
where the perturbative Green's function
technique was employed. Should such nonanalytic behavior of the
scattering amplitude continues to be exhibited
in a relativistic field theory it would be a matter of concern.
We have mentioned that analyticity and causality are
closely related while deriving results from axiomatic field theories.
The analyticity and crossing properties of
scattering amplitude were investigated, for $D>4$,
 in the LSZ formulation only recently.
We summarize the essential conclusions of \cite{jmjmp} which
will be utilized in the study of analyticity of scattering amplitudes
in compactified theories.
     It was shown, in the LSZ formalism,  that
the scattering amplitude has desire attributes in the following sense:
(i)  We proved the generalization of the
Jost-Lehmann-Dyson theorem for the retarded function \cite{jl,dyson}
for the $D>4$  case \cite{jmplb}. (ii) Subsequently, we
  showed the existence of the
Lehmann-Martin ellipse for such a theory. (iii)
Thus a dispersion relation can be written down in $s$ for fixed $t$
when the momentum transfer squared lies inside Lehmann-Martin ellipse
\cite{leh2,martin1}. (iv) The analog of Martin's theorem can be
derived in the sense that the scattering amplitude is analytic the
product domain $D_s\otimes D_t$ where $D_s$ is the
cut $s$-plane and $D_t$ is a domain in the $t$-plane such that the scattering
amplitude is analytic inside a disk, $|t|<{\tilde R}$, ${\tilde R}$ is radius
of the
disk and it is  independent of
$s$.  Thus the partial wave expansion converges inside this bigger domain.
(v) We also derived the analog of Jin-Martin \cite{jm}upper bound on
the scattering amplitude which states
that the fixed $t$ dispersion relation in $s$ does not require more than
two subtractions. (vi) Consequently, a generalized
Froissart-Martin bound was be proved.\\
In order to accomplish our goal for a $D=4$ theory which arises from $S^1$
compactification of a $D=5$ theory i.e. to prove nonforward
dispersion relations, we have to establish the results (i) to (iv) for
this theory. It is important to point out, at this  juncture, that (to be
elaborated in the sequel)
the spectrum of the theory consists of a massive particle of the original
five dimensional theory and a tower of Kaluza-Klein states.
Thus the requisite results (i)-(iv) are to obtained in this context in
contrast to the results of the D-dimensional theory with a single massive
neutral scalar
field. \\
The developments in this case are similar
to the ones derived for $D=4$ theories. However, certain subtle issues
had to be surmounted in order to
prove analyticity and crossing properties for theories defined in higher
dimensions, $D>4$. The author was drawn into
the topic through discussions with
Andr\'e Martin (Martin private discussions). He expressed his concern that
if the analyticity would be violated in a compactified field theory
then several rigorous results derived from axiomatic
field theories will be questioned. In particular, what would be the fate
of Froissart-Martin bound for such a theory?
The author undertook the study of analyticity of scattering amplitude in
a field theory with a compact spatial dimension.
It is necessary to start from fundamental axioms of LSZ for an
uncompactified field theory in higher spacetime
dimension and compactify a spatial dimension and examine the analyticity
properties of the scattering amplitude. This is
the topic to be discussed in this article.\\
As mentioned earlier, the analyticity property of the amplitude in
nonrelativistic potential scattering has been
investigated long ago. The result of Khuri was that scattering amplitude
for a potential with a compact coordinate
violates analyticity was a surprise. However, we should carefully analyze
the implications of Khuri's result.
We recall that in QFT the analyticity of an amplitude and causality are
intimately related. The relativistic
invariance of the theory implies that no signal can travel faster than
the velocity of light. Therefore, two local
(bosonic) operators commute when they are separated by spacelike distance.
 As we shall discuss later, the
key ingredient to prove analyticity of the amplitude is the axiom of
microcausality. In the context of nonrelativistic
potential scatterings, the theory is invariant under Galilean transformations.
 Consequently, the concept of microcausality
is not envisaged in potential scattering. Therefore, the lack of analyticity
of an amplitude, in certain cases, is not
so serious an issue as would be the case if analyticity is not respected in
a relativistic QFT. We shall proceed while
keeping in mind the preceding remarks.\\
The article is organized as follows.  In the next section, Section 2,
we present a very brief account of
 Khuri's results to familiarize the reader with his
formulation of the problem for a potential which has a compact
spatial dimension.
The third section, Section 3, is devoted to a short review of LSZ formalism.
We present the LSZ reduction technique for a massive
neutral scalar field theory in higher dimensions, i.e 
in five spacetime dimensions, $D=5$.
All the requisite ingredients to prove dispersion relations
are summarized here. We briefly discuss crossing symmetry and touch
upon derivation of the Lehmann
ellipses. We need these two results to write down dispersion relations.
Next we discuss the $S^1$ compactification of the flat space five dimensional
theory. The $R^{4,1}$ manifold is compactified to $R^{3,1}\otimes S^1$.
 The starting point is to consider
a single massive scalar field theory
defined in a flat five dimensional manifold, $R^{4,1}$. Thus there is
one massive scalar of mass $m_0$ living  in $D=5$.
When we compactify one spatial coordinate on $S^1$, the resulting theory
defined on the manifold $R^{3,1}\otimes S^1$
is endowed with the following spectrum. There is a massive scalar of
mass $m_0$. In addition, there is a tower of KK states
whose mass spectrum is $m_n^2=({n\over R})^2$ where R is the compactification
radius and $n\in {\bf Z}$. In fact each KK
state is endowed with an integer KK charge, $q_n\in Z$ which is conserved.
 Therefore, the compactified QFT has various
features which differ from a nonrelativistic potential model.
 The next section, Section 4,
 is devoted to investigate
analyticity properties of the scattering amplitude for the theory alluded
to above. We systematically derive the
spectral representation for the absorptive amplitudes. Then discuss the
crossing properties. We touch upon the
Jost-Lehmann-Dyson theorem for this case which has been derived for
a field theory defined in higher
dimensions, $D>4$. However, it is essential to consider the existence
of Lehmann ellipses. The proof of dispersion
relation requires the existence of Lehmann ellipses, especially,
the Large Lehmann Ellipse (LLE). Subsequently,
we write down the unsubtracted, fixed-t dispersion relations. In fact,
the elastic scattering amplitude for $n=1$ states
is considered. It is shown that there is no violation of analyticity
in this case. Indeed, our proof goes beyond the results
of Khuri since we have proven the nonforward dispersion relations.
We derive a few corollaries based on our main
results.
\\
Section 5 is devoted to prove the generalized unitarity relation in
the LSZ formulation for the theory under considerations.  
It has two purpose. First, we note that the unitarity constraint
already provides a preview of crossing as will be discussed. We have
not proved crossing explicitly since it is not our main goal. The
second important result is that the unitarity of S-matrix implies
that only a finite number KK excited states contribute to the
spectral representations as physical intermediate states. We
draw attention of the reader to a very important observation that
only the physical states appear as intermediate states in the
spectral representation. It is unitarity, the nonlinear relation,
which cuts off the sum to a finite number of terms when we sum 
over the KK towers. Notice that when we derive the spectral
representation for the matrix element of the causal commutator
of the source currents the sum over intermediate states is
the entire KK tower \footnote{ I thank Luis Alvarez Gaume for raising
this question}. There is no way to conclude, in the linear program,
that the sum could be over finite number of KK states. \\
 We had proposed another novel way \cite{jmjmp15} to look for the evidence
large-radius-compactification proposal. In Section 6 we proceed to examine that
idea. 
We  argue  
\cite{nm} that precise measurement of $\sigma_t$ at LHC energy and beyond
might provide a clue to look for evidence for  the 
large-radius-compactification hypothesis. If a theory is defined
in higher dimensional flat space, $D>4$, then the Froissart bound on $\sigma_t$
is modified. The proof is derived from LSZ axioms \cite{jmjmp}.
Suppose, one extra
dimension is decompactified at LHC energies and the total cross section has
an energy dependence which violates the $D=4$ Froissart-Martin bound. In the
light of above remark, one should conclude immediately that some of the axioms
of local field theories might be violated. Instead, we should interpret
the observed
energy dependence a signal of decompactification of extra dimensions.
We have analyzed
the data \cite{nm} recently. However, we conclude that there is no
conclusive evidence
for violation of the Froissart-Martin bound. We feel that more precise
measurements of
$\sigma_t$ might provide some hints on this issue.

\bigskip

\noindent {\bf 2. Non-relativistic Potential Scattering for 
$R^3\otimes S^1$ Geometry}

\bigskip

\noindent In this section, we shall discuss the essential results of
Khri \cite{khuri1} where he considered a nonrelativstic potential
scattering. He introduced a spherically symmetric potential in
three dimensions with an additional compact coordinate.
Let us consider the set up for potential   scattering in the framework of 
nonrelativistic quantum mechanics. The potential, $V(r)$,
is spherically symmetric, where $r=|{\bf r}|$. It is chosen to
be a short range potential with good behaviors for large $r$,
see \cite{khuri2} for
details. The starting point is the Schr\"odinger equation
 \bea
 \label{kh1a}
 \bigg[{ \nabla}^2+k^2- V(r) \bigg]\Psi({\bf r})=0
\eea
This equation is expressed in the dimensionless form so that the mass
and Planck's constant do not appear. The solution
to the above equation is chosen such that for large $r$ there is a
plane wave part and an outgoing spherical wave component. Therefore,
\bea
\label{kh1b}
\Psi({\bf r})=e^{i{\bf k}.{\bf r}}+
\int d^3r'G_0({\bf r}-{\bf r}')V({\bf r}')\Psi({\bf r}')
\eea
with the free plane wave solution
${1\over{2\pi^2}}e^{i{\bf k}.{\bf x}}$ and the Green's function
$G_0({\bf r}-{\bf r}')$ satisfies the free Schr\"odinger equation
\bea
\label{kh1c}
 \bigg[{ \nabla}^2+k^2  \bigg]  G_0({\bf r}-{\bf r}') =
\delta^3({\bf r}-{\bf r}')
 \eea
 The solution satisfies the desired boundary conditions.
The asymptotic form of the above solution (\ref{kh1b}) is
 \bea
 \label{kh1d}
 \Psi({\bf r})\approx e^{i{\bf k}.{\bf r}}+{1\over{|{\bf r}|}}
e^{i{\bf | k|}|{\bf r}|}f(k,cos\theta)
 \eea
 Here $r=|{\bf r}|$ and $cos\theta$ is the center of mass scattering angle.
 The coefficient of the spherical  wave component, $f(k,cos\theta)$ is
 defined to be the scattering amplitude. This is the Born amplitude
 and we iterate this procedure to get the higher order correction.
 Moreover, Khuri \cite{khuri2} went through a rigorous procedure to
 study the analyticity of the scattering amplitude.  He proved that
 the scattering amplitude is analytic in the upper half $k$-plane
 for fixed $cos\theta$. Furthermore, it is bounded in the upper half plane
 and also on the real axis for a general class of potentials
 which have good convergent behavior as $r\rightarrow \infty$. It was an
 important result at that juncture. It was also a surprising and unexpected
 outcome since nonrelativistic  theories are not endowed
with principle of microcausality as is the
 case in the relativistic theories. The velocity of light is the
 limiting velocity for latter theories. Therefore, analyticity and
 causality are intimately connected only in relativistic theories.  \\
Now we turn attention to Khuri's study of analyticity of scattering
 amplitude in a nonrelativistic theory in three spatial dimension which
also has one compactified spatial coordinate.
  Khuri \cite{khuri1}, in 1995,   envisaged scattering of a particle in
a space with $R^3\otimes S^1$ topology. We  provide a brief account
of his work and incorporate  his important conclusions.
 We refer  to the original paper to the  interested reader.
 The notations of \cite{khuri1}
will be followed.
The compactified coordinate is a circle of radius $R$ and it
 is assumed that
the radius is {\it small} i.e. ${1\over R}>> 1$ where dimensionless
units were used. We mention here that the five dimensional theory is
  defined in a
flat Minkowski space. The only mass scale available to
us is the mass of the
particle; therefore, ${1\over R}$  is much larger than this scale.
The potential, $V(r,  \Phi)$, is such that it is periodic in the angular
 coordinate, $ \Phi$, of $S^1$; $\bf r\in R^3$ and
$r=|{\bf r}|$. The potential, $V(r,\Phi)$,  belongs to a broad class such
 that for large $r$  this class of
potentials fall off like  $e^{-\mu r}/r$ as $r\rightarrow \infty$.
  Moreover,  $V(r,\Phi)=V(r,\Phi+2\pi)$. The scattering amplitude depends
 on three variables - the momentum
of the particle, $k$, the scattering angle $\theta$, and an integer
$n$ which appears due to the periodicity of the $\Phi$-coordinate.
Thus forward scattering amplitude is denoted by $T_{nn}(K)$,
 where $K^2=k^2+{{n^2}\over{R^2}}$.
The starting point is the Schr\" odinger equation
\bea
\label{kh1}
\bigg[{ \nabla}^2+{{1\over R^2}}{{\partial}^2\over{\partial\bf\Phi}^2}
+K^2-V(r,\Phi )\bigg]\Psi({\bf r},\Phi)=0
\eea
The free plane wave solutions are
\bea
\label{kh2}
\Psi_0({\bf x},\Phi})={{1\over{(2\pi)^2}}e^{i{\bf k}.{\bf x}}e^{i n\Phi}
\eea
and $n\in {\bf Z}$.
The total energy is defined to be
\bea
\label{kh3}
{\bf K}^2=k^2+{{n^2}\over{R^2}}
\eea
The free Green's function (in the presence of a compact coordinate) assumes
the following form
\bea
\label{kh4}
G_0({\bf K};{\bf x},\Phi:{\bf x'},\Phi ')=-{{1\over{(2\pi)^4}}}
\sum_{n=-\infty}^{n=+\infty}\int d^3p{{e^{i{\bf p}.({\bf x}-{\bf x'})}
e^{in.(\Phi-\Phi')}}
\over{[p^2+{{n^2}\over{R^2}}-{\bf K}^2-i\epsilon]}}
\eea
The free Green's function satisfies the free Schr\" odinger equation
\bea
\label{kh5}
\bigg[{ \nabla}^2+{{1\over R^2}}{{\partial}^2\over{\partial\bf\Phi}^2}
+K^2\bigg] G_0({\bf K};{\bf x},\Phi:{\bf x'},\Phi ')
=\delta^3({\bf x}-{\bf x'})\delta(\Phi -\Phi')
\eea
The $d^3p$ integration can be performed in the expression
(\ref{kh4}) leading to
\bea
\label{kh6}
G_0({\bf K};{\bf x} -{\bf x'};\Phi-\Phi ')=-{{1\over{(8\pi^2)}}}
\sum_{n=-\infty}^{n=+\infty}{{ e^{i{\sqrt{K^2-{(n^2/{R^2})}}|{\bf x}-{\bf x'}|}}}\over{|{\bf x}-{\bf x'}|}}
e^{in(\Phi-\Phi')}
\eea
Khuri introduced the following prescription for ${\sqrt{K^2-{n^2/{R^2}}}}$. It
is defined in such a way that when  $n^2/{R^2}>K^2$
\bea
\label{kh7}
i{\sqrt{K^2-n^2/{R^2}}}\rightarrow -{\sqrt{n^2/{R^2}-K^2}},~~n^2>K^2R^2
\eea
Note that the series expansion for
$G_0({\bf K};{\bf x} -{\bf x'};\Phi-\Phi ')$ as expressed in
(\ref{kh6}) is strongly damped for large enough $|n|$. A careful analysis,
as was carried
out in ref. \cite{khuri1},   shows that the Green's function is well
defined and bounded, except for $|{\bf x}-{\bf x'}|\rightarrow 0$;  similar
to the properties of Green's functions in potential scattering for a
fixed $K^2$.  Khuri \cite{khuri2}  expressed
 the scattering integral equation for the potential $V(r,\Phi)$ as
\bea
\label{kh8}
\Psi_{k,n}({\bf x},\Phi)
=e^{i{\bf k}.{\bf x}}e^{in\Phi}+
\int_0^{2\pi} d\Phi'\int d^3{\bf x'}G_0({\bf K};|{\bf x} -{\bf x'}|;|\Phi-\Phi '|)V({\bf x'},\Phi')\Psi_{k,n}({\bf x'},\Phi')
\eea
The expression for the scattering amplitude is extracted
from the large $|{\bf x}|$ limit when one looks at the asymptotic behavior of
the wave function,
\bea
\label{kh9}
\Psi_{{\bf k},n}\rightarrow e^{{\bf k}.{\bf x}}e^{in\Phi}+ \sum_{m=-[KR]}^{+[KR]}T({\bf k'},m:{\bf k},n){{e^{ik'_{mn}|{\bf x}|}\over{|{\bf x}|}}}e^{im\Phi}
\eea
where $[KR]$ is the largest integer less than $KR$ and
\bea
\label{kh10}
k'_{mn}={\sqrt{k^2+{{n^2}\over{R^2}}-{{m^2}\over{R^2}}}}
\eea
He also identifies a conservation rule:
$K^2=k^2+(n^2/{R^2})=k'^2+(m^2/{R^2})$. Thus it is argued that
 that the scattered wave has only $(2[KR]+1)$ components and those states with
$(m^2/(R^2)>k^2+(n^2/{R^2}))$ are exponentially damped for large
$|{\bf x}|$ and consequently,  these do not appear in the scattered wave
 (see eq. (\ref{kh7})).
Now the scattering amplitude is extracted from equations (\ref{kh8})
and (\ref{kh9}) to be
\bea
\label{kh11}
T({\bf k'},n';{\bf k},n)=-{{1\over{8\pi^2}}}\int d^3{\bf x'}\int_0^{2\pi} d\Phi'e^{-i{\bf k'}.{\bf x'}}e^{-in'\Phi'}V({\bf x'},\Phi')\Psi_{{\bf k},n}({\bf x'},\Phi')
\eea
The condition,  $k'^2+n'^2/{R^2}=k^2+n^2/{R^2}$ is to be satisfied.
Thus the scattering amplitude describes the process where incoming wave
$|{\bf k},n>$ is scattered to final state $|{\bf k'},n'>$.\\
Remark: Reader should pay attention to the expression for the discussion
of scattering processes in relativistic QFT in the sequel and note
the similarities and differences as discussed in subsequent   sections.  \\
Formally, the amplitude assumes the following form for the full Green's
 function
\bea
\label{kh12}
T({\bf k'},n';{\bf k},n)-T_B=&&-{{1\over{8\pi^2}}}\int ....\int d^3{\bf x}d^3{\bf x'} d\Phi d\Phi'e^{-i({\bf k'}.{\bf x'}+n'\Phi')}
V({\bf x'},\Phi')\nonumber\\&&G({\bf K};{\bf x'},{\bf x};\Phi',\Phi) V({\bf x},\Phi)e^{i({\bf k}.{\bf x}+n\Phi)}
\eea
Here $T_B$ is the Born term.
\bea
\label{kh12a}
T_B=-{1\over{8\pi^2}}\int d^3x\int_0^{2\pi}d\Phi
e^{i({\bf k}-{\bf k}).{\bf x}}V(x,\Phi)e^{i(n-n')\Phi}
\eea
 Full Green's function satisfied an equation with the full Hamiltonian
\bea
\label{kh13}
\bigg[{ \nabla}^2+{{1\over R^2}}{{\partial}^2\over{\partial\bf\Phi}^2} +K^2-V({\bf x},\Phi) \bigg]G({\bf K};{\bf x},{\bf x'},\Phi,\Phi' )=\delta^3({\bf x}-{\bf x'})\delta(\Phi-\Phi')
\eea
This is the starting point of computing scattering amplitude
perturbatively in potential scattering \cite{gw}. Khuri \cite{khuri1}
proceeds to study the analyticity
properties of the amplitude and it is a parallel development similar to
investigations done in the past. In the context of a theory with compact space
dimension he analysed the amplitude  $T_{nn}(K)$ to the second order in
the Born approximation for $n=1$.\\
Khuri \cite{khuri1}  explicitly computed the second born term $T^{(2)}$
for the forward amplitude, for the choice $n=1$. He has discovered that
the analyticity of the forward amplitude
breaks down with a counter example; where $T_{nn}(k)$ does not
satisfy dispersion relations for a class of Yukawa-type
potentials of the form
\bea
\label{kh14}
V(r,\Phi)=u_0(r)+2\sum_{m=1}^N u_m(r)cos({m\Phi})
\eea
  where $u_m(r)=\lambda_m{{e^{-\mu r}} \over{ r}}$. Khuri noted an
 important feature of his studies  that in the case when scattering
 theory was applied
  perturbatively in $R^3$ space the resulting amplitude satisfied
 analyticity properties for similar Yukawa-type
 potentials. Thus there has been concerns\footnote{Andre Martin brought
 the work of Khuri \cite{khuri1} to my attention and persuaded me to
undertake this
  investigation.}  when non-analyticity of the aforementioned
scattering amplitude was discovered in
the non-relativistic quantum mechanics by Khuri
  in the space with the topology $R^3\otimes S^1$.\\
  We shall describe the framework of our investigation in the next section.
 We remark in passing that the analyticity of scattering amplitude
 in nonrelativitic
  scattering is not such a  profound property as in the relativistic QFT
although the analyticity in non-relativistic potential scattering has been
investigated
  quite thoroughly in the past \cite{gw}. However, it is to be noted that
 in absence a limiting velocity (in relativistic case velocity of
light, c, profoundly
  influences the study of the analyticity of amplitudes)
 the microcausality is not enforced in nonrelatvistic processes.
 As we shall show (and has been
  emphasized in many classic books) there is indeed a deep
connection between microcausality and analyticity. When a spatial
 dimension is
  compactified on $S^1$, the coordinate on the circle is periodic;
we can understand  microcausality as follows.
The compact coordinate $y$ is periodic. Therefore, we can define
spacelike separation between two points.  We keep this aspect in mind and
 we shall undertake a systematic study of the analyticity of scattering
amplitude in the  sequel.

\bigskip

\noindent{\bf 3.  Quantum Field Theory with Compact Spatial Dimensions}

\bigskip

\noindent First, we present the LSZ formalism for a $D=5$ massive theory
in the  flat Minkowski spacetime. Subsequently, we discuss the $S^1$
compactification in detail. 

\bigskip

\noindent {\bf 3.1 Quantum Field Theory in $D=5$ Spacetime}.

\bigskip
\noindent
We have shown in \cite{jm1} (henceforth referred as I)
 that the forward scattering amplitude of a theory, 
defined on the manifold $R^{3,1}\otimes S^1$,  satisfied
dispersion relations. This result was obtained in the frame works of 
the LSZ formalism. Thus the proof of the forward dispersion relations
will not be presented in this review. The interested reader can get
the details in I. We summarize, in this
subsection, the  starting points of I as stated below.
\\
 We  considered a neutral, scalar field theory with mass $m_0$
 in flat five dimensional Minkowski space $R^{4,1}$. 
It is assumed that the particle is stable and there are no bound states. 
The notation is  that the  spacetime coordinates
are, $\hat x$,
and all operators are denoted with a {\it hat} when they are defined in 
the five dimensional  space where the spatial coordinates 
are noncompact.The  LSZ axioms are \cite{lsz}:\\
{\bf A1.} The states of the system are represented in  a
Hilbert space, ${\hat{ \cal H}}$. All the physical observables are self-adjoint
operators in the Hilbert space, ${\hat{\cal H}}$.\\
{\bf A2.} The theory is invariant under inhomogeneous Lorentz 
transformations.\\
{\bf A3.} The energy-momentum of the states are defined. It follows from the
requirements of  Lorentz  and translation invariance that
we can construct a representation of the
orthochronous  Lorentz group. The representation
corresponds to unitary operators, ${\hat U}({\hat a},{\hat \Lambda})$,  
and the theory is invariant 
under these transformations. Thus there are Hermitian operators corresponding
to spacetime translations, denoted as ${\hat P}_{{\hat\mu}}$, with 
${\hat \mu}=0,1,2,3,4$ which have following
properties:
\bea
\bigg[{\hat P}_{\hat\mu}, {\hat P}_{\hat\nu} \bigg]=0
\eea
If ${{\hat{\cal F}}({\hat x})}$ is any Heisenberg operator then its 
commutator with ${\hat P}_{\hat\mu}$
is
\bea
\bigg[{\hat P}_{\hat\mu}, {\hat{\cal F}}({\hat x}) \bigg]=
i{\hat\partial}_{\hat\mu} {\hat{\cal F}}({\hat x})
\eea
It is assumed that the operator does not explicitly depend on spacetime 
coordinates.
  If we choose a representation where the translation operators, 
${\hat P}_{\hat\mu}$,
are diagonal and the basis vectors $|{\hat p},\hat\alpha>$  span the Hilbert 
space,
${\hat{\cal H}}$, 
\bea
{\hat P}_{\hat\mu}|{\hat p},{\hat\alpha}>=
{\hat p}_{\hat\mu}|{\hat p},\hat\alpha>
\eea
then we are in a position to make more precise statements: \\
${\bullet}$ Existence of the vacuum: there is a unique invariant vacuum state
$|0>$ which has the property
\bea
{\hat U}({\hat a},{\hat\Lambda})|0>=|0>
\eea
The vacuum is unique and is Poincar\'e invariant.\\
${\bullet}$ The eigenvalue of ${\hat P}_{\hat\mu}$, ${\hat p}_{\hat\mu}$,  
is light-like, with ${\hat p}_0>0$.
We are concerned  only with  massive stated in this discussion. If we 
implement
infinitesimal Poincar\'e transformation on the vacuum state then
\bea
{\hat P}_{\hat\mu}|0>=0,~~~ {\rm and}~~~ {\hat M}_{\hat{\mu}\hat\nu}|0>=0
\eea
from above postulates and note that ${\hat M}_{{\hat\mu}\hat\nu}$ are 
the generators of Lorentz
transformations.\\
{\bf A4.} The locality of theory implies that a (bosonic) local operator 
at spacetime point
${\hat x}^{\hat\mu}$ commutes with another (bosonic) 
local operator at ${\hat x}'^{\hat\mu}$ when  their
separation is spacelike i.e. if $({\hat x}-{\hat x}')^2<0$. 
Our Minkowski metric convention
is as follows: the inner product of two 5-vectors is given by
${\hat x}.{\hat y}=
{\hat x}^0{\hat y}^0-{\hat x}^1{\hat y}^1-...-{\hat x}^{4}{\hat y}^{4}$.
Since we are dealing with a neutral scalar
field, for the field operator ${\hat\phi }({\hat x})$: 
${{\hat\phi}({\hat x})}^{\dagger}={\hat\phi}({\hat x})$ i.e.
 ${\hat\phi} ({\hat x})$ is Hermitian.
By definition it  transforms as a scalar under inhomogeneous Lorentz
transformations 
\bea
 {\hat U}({\hat a},{\hat\Lambda}){\hat\phi}({\hat x}){\hat U}({\hat a},
{\hat\Lambda})^{-1}={\hat\phi}({\hat\Lambda} {\hat x}+{\hat a})
\eea
The micro causality, for two local field operators,  is stated to be 
\bea
\bigg[{\hat\phi}({\hat x}),{\hat\phi}({\hat x}') \bigg]=0,~~~~~for~~
({\hat x}-{\hat x}')^2<0
\eea
It is well known that, in the LSZ formalism,  we are concerned with vacuum
expectation values of time ordered products of operators as well as
with the  the retarded product of fields. The requirements of the above 
listed axioms
lead to certain relationship, for example, between vacuum expectation values 
of
R-products of operators. Such a set of relations are termed as the 
{\it linear relations} and the importance of
the above listed axioms is manifested through these relations.
 In contrast, unitarity
imposes {\it nonlinear}  constraints on amplitude. For example,
 if we expand an amplitude
in partial waves, unitarity demands  certain positivity conditions
 to be satisfied by
the partial wave amplitudes. \\
We summarize below some of the important aspects of LSZ formalism 
as we utilize them
through out the present investigation. Moreover, the conventions 
and definitions of I will be followed
for the conveniences of the reader.\\
(i) The asymptotic condition: According to LSZ the field theory accounts
 for the asymptotic observables.
These correspond to particles of definite mass, charge and spin etc.
 ${\hat \phi}^{in}({\hat x})$ represents
the free field in the remote past. A  Fock space is generated 
by the field operator. 
The physical observable can be
expressed in terms of these fields.\\
(ii)  ${\hat\phi}( \hat x)$ is the interacting field. LSZ  technique 
incorporates a prescription to relate the
interacting field, ${\hat\phi}( \hat x)$, with ${\hat\phi}^{in}({\hat x})$; consequently, the asymptotic fields are
defined with a suitable limiting procedure. Thus we introduce the notion 
of the adiabatic switching off
of the interaction. A cutoff adiabatic function is postulated such that 
this function controls the 
interactions. It is $\bf 1$ at finite interval of time and it has a smooth 
limit of passing to zero
as $|t| \rightarrow \infty$. It is argued that when adiabatic switching 
is removed we can define
the physical observables.\\
(iii) The fields  ${\hat \phi}^{in}({\hat x})$ and 
${\hat\phi}( \hat x)$ are related as follows:
\bea
\label{z}
{\hat x}_0\rightarrow -\infty~~~~{\hat\phi}({\hat x})
\rightarrow {\hat Z}^{1/2}{\hat\phi}^{in}({\hat x})
\eea
By the first postulate, ${\hat\phi}^{in}({\hat x})$ creates 
free particle states. However,
in general ${\hat\phi}({\hat x})$ will create 
multi particle states besides the single
particle one since it is the interacting field. Moreover, 
$<1|{\hat\phi}^{in}({\hat x})|0>$ 
and
 $<1|{\hat\phi}({\hat x})|0>$ carry same functional dependence in 
$\hat x$.  If the factor 
of $\hat Z$ were not the scaling relation between the two
fields (\ref{z}), then canonical commutation relation for each of the 
two fields ( i.e. ${\hat\phi}^{in}({\hat x})$ and  ${\hat\phi}({\hat x})$) 
 will be the same.
Thus in the absence of $\hat Z$ the two theories will be identical. 
Moreover, the
postulate of asymptotic condition states that in the remote future
\bea
{\hat x}_0\rightarrow \infty~~~~{\hat\phi}({\hat x})
\rightarrow {\hat Z}^{1/2}{\hat\phi}^{out}({\hat x}).
\eea
We may as well construct a Fock space utilizing 
${\hat\phi}^{out}({\hat x})$ as we could   with ${\hat\phi}({\hat x})^{in}$.
Furthermore, the vacuum is unique for ${\hat\phi}^{in}$,  
${\hat\phi}^{out}$ and ${\hat\phi}({\hat x})$. The
normalizable single particle states are the same i.e.
${\hat\phi}^{in}|0>={\hat\phi}^{out}|0>$. We do not display 
${\hat Z}$  from now on. If at all
any need arises,  ${\hat Z}$ can be introduced in the relevant expressions.\\
We define creation and annihilation operators for ${\hat\phi}^{in}$,
${\hat\phi}^{out}$. We recall that   ${\hat\phi}({\hat x})$  is not a 
free field. Whereas the fields ${\hat\phi}^{in,out}({\hat x})$ satisfy 
the free field 
equations $[{\Box}_5+m^2_0]{\hat\phi}^{in,out}({\hat x})=0$, 
the interacting  field satisfies an equation of motion
which is endowed with  a source current: 
$[{\Box}_5+m^2_0]{\hat\phi}({\hat x})={\hat j}({\hat x})$.
 We  may use  the plane wave basis for simplicity in certain 
computations; however,
in a more formal approach, it is desirable to use wave packets.\\
The relevant vacuum expectation values of the products of operators in 
LSZ formalism are either the time ordered
products (the T-products) or the retarded products (the R-products). 
We shall mostly use the R-products and 
we use them extensively  throughout this investigation. 
It is defined as 
\bea
R~{\hat\phi}({\hat x}){\hat\phi}_1({\hat x}_1)...{\hat\phi}_n({\hat x}_n)
=&&(-1)^n\sum_P\theta({\hat x}_0-{\hat x}_{10})
\theta({\hat x}_{10}-{\hat x}_{20})...
\theta({\hat x}_{n-10}-{\hat x}_{n0})\nonumber\\&&
[[...[{\hat\phi}({\hat x}),{\hat\phi}_{i_1}
({\hat x}_{i_1})],{\hat\phi}_{i_2}({\hat x}_{i_2})]..],{\hat\phi}_{i_n}
({\hat x}_{i_n})]
\eea
note that $R{\hat\phi}({\hat x})={\hat\phi}({\hat x})$ and  P 
stands for all the permutations ${i_1,....i_n}$  of  $1,2...n$.
The R-product is hermitian for hermitian fields ${\hat\phi}_i({\hat x}_i)$
 and
the product is symmetric under exchange of any fields
${\hat\phi}_1({\hat x}_1)...{\hat\phi}_n({\hat x}_n)$. Notice that the 
field ${\hat\phi}({\hat x})$ is kept where it is
located in  its position.
We list below some of the important properties of the $R$-product for 
future use \cite{fr1}:\\
(i) $R~{\hat\phi}({\hat x}){\hat\phi}_1({\hat x}_1)...{\hat\phi}_n({\hat x}_n) 
\ne 0$ only if
${\hat x}_0>~{\rm max}~\{{\hat x}_{10},..{\hat.x}_{n0} \}$.\\
(ii) Another important property of the R-product is that
\bea
R~{\hat\phi}({\hat x}){\hat\phi}_1({\hat x}_1)...{\hat\phi}_n({\hat x}_n) = 0
\eea
whenever the time component ${\hat x}_0$, appearing in the argument of ${\hat\phi}({\hat x})$ whose
position is held fix, is less than time component of any of the four vectors
$({\hat x}_1,...{\hat x}_n)$ 
appearing in the arguments of 
${\hat\phi}({\hat x}_1)...{\hat\phi}({\hat x}_n)$.\\
(iii) We recall that
\bea
{\hat\phi}({\hat x}_i)\rightarrow {\hat\phi}({\hat\Lambda} {\hat x}_i)={\hat U}({\hat\Lambda},0){\hat\phi}({\hat x}_i){\hat U}({\hat\Lambda},0)^{-1}
\eea
Under Lorentz transformation ${\hat U}({\hat\Lambda},0)$. Therefore,
\bea
R~{\hat\phi}({\hat\Lambda}{\hat x}){\hat\phi}({\hat\Lambda}{\hat x}_i)...
{\hat\phi}_n({\hat\Lambda} {\hat x}_n)={\hat U}({\hat\Lambda},0)
R~\phi(x)\phi_1(x_1)...\phi_n(x_n)U(\Lambda,0)^{-1}
\eea
And
\bea
 {\hat\phi}_i({\hat x}_i)\rightarrow{\hat\phi}_i({\hat x}_i+{\hat a})=
e^{i{\hat a}.{\hat P}}{\hat\phi}_i({\hat x}_i)e^{-i{\hat a}.{\hat P}}
\eea
 under spacetime translations. Consequently,
\bea
R~{\hat\phi}( {\hat x}+{\hat a}){\hat\phi}({\hat x}_i+{\hat a})
...{\hat\phi}_n({\hat x}_n+{\hat a})=
e^{i{\hat a}.{\hat P}}R~{\hat\phi}({\hat x}){\hat\phi}_1({\hat x}_1)...
{\hat\phi}_n({\hat x}_n)e^{-i{\hat a}.{\hat P}}
\eea
 Therefore,   the vacuum expectation value of the R-product
dependents only on  difference between pair of coordinates: in other words it
depends on the
following set of coordinate differences: 
${\hat\xi}_1
={\hat x}_1-{\hat x},{\hat\xi}_2={\hat x}_2-{\hat x}_1...
{\hat\xi}_n={\hat x}_{n-1} -{\hat x}_n$ as a consequence of
translational invariance. \\
(iv) The retarded property of R-function and the asymptotic conditions lead
 to the following relations.
\bea
[R~{\hat\phi}({\hat x}){\hat\phi}_1({\hat x}_1)...
{\hat\phi}_n({\hat x}_n),{\hat\phi}^{in}_l({\hat y}_l)]=&&
i\int d^5{\hat y}'_l\Delta({\hat y}_l-{\hat y'_l})({\Box}_{{5\hat y}'}+
{\hat m}_l^2)\times \nonumber\\&&
R~{\hat\phi}({\hat x}){\hat\phi}_1({\hat x}_1)...
{\hat\phi}_n({\hat x}_n){\hat\phi}_l({\hat y}'_l)
\eea
Note: here ${\hat m}_l$ stands for the mass of a field in five dimensions.
We may define 'in' and 'out' states in terms of the creation operators 
associated with 'in' and 'out' fields as follows
\bea
\label{fock1}
|{\hat k}_1,{\hat k}_2,....{\hat k}_n~in>=
{\hat a}_{in}^{\dagger}({\hat{\bf k}}_1){\hat a}_{in}^{\dagger}
({\hat{\bf k}}_2)...
{\hat a}_{in}^{\dagger}({\hat{\bf k}}_n)|0>
\eea
\bea
\label{fock2}
|{\hat k}_1,{\hat k}_2,....{\hat k}_n~out>={
\hat a}_{out}^{\dagger}({\hat{\bf k}}_1){\hat a}_{out}^{\dagger}
({\hat{\bf k}}_2)...
{\hat a}_{out}^{\dagger}({\hat{\bf k}}_n)|0>
\eea
We can construct a complete set of states either starting from 'in'  
field operators or the 'out' field operators and each complete set
will span the Hilbert space,  ${\hat{\cal H}}$. Therefore, a unitary 
operator will relate the two sets of states in this Hilbert
space. This is a heuristic way of introducing the concept of the $S$-matrix.
 We shall define $S$-matrix elements
through LSZ reduction technique in subsequent section.\\
We shall not distinguish between notations like 
${\hat\phi}^{out,in}$ or ${\hat\phi}_{out,in}$ and therefore, 
there might be use
of the sloppy notation in this regard.\\
We record the following important remark {\it en passant}. 
The generic matrix element 
$<{\hat\alpha}|{\hat\phi}({\hat x}_1){\hat\phi}({\hat x}_2)...|{\hat\beta}>$
is not an ordinary function but a distribution. Thus it is to be always
understood as smeared with a Schwartz type test function $f\in {\cal S}$. 
The test
function is infinitely differentiable and it goes to zero along with all its
derivatives faster than any power of its argument. We shall formally derive 
expressions
for scattering amplitudes and the absorptive parts by employing the 
LSZ technique. It is to be understood that
these are generalized functions and such matrix elements are properly defined
with smeared out test functions.\\
We  obtain below the expression for the K\"allen-Lehmann representation for the
five dimensional theory. It will help us to transparently expose, 
as we shall recall in the next section, the consequences of
$S^1$ compactification. Let  us consider the vacuum expectation value 
(VEV) of the
commutator of two fields in the $D=5$ theory: 
$<0|[{\hat\phi}({\hat x}), {\hat\phi}({\hat y})]|0>$. We
introduce a complete set of states between product of the fields after 
opening up the commutator. Thus
we arrive at the following expression by adopting the standard arguments,
\bea
\label{KL}
  <0|[{\hat\phi}({\hat x}), {\hat\phi}({\hat y})]|0>=
\sum_{\hat\alpha}\bigg(<0|{\hat\phi}(0){\hat\alpha}>e^{-i{\hat p}_{\hat\alpha}.
({\hat x}-{\hat y})}
<{\hat\alpha}|{\hat\phi}(0)|0>-({\hat x}{\leftrightarrow}{\hat y}) \bigg)
\eea
Let us define
\bea
{\hat\rho}({\hat q})=(2\pi)^4\sum_{\hat\alpha}
\delta^5({\hat q}-{\hat p}_{\hat\alpha})|<0|{\hat\phi}(0)|{\hat\alpha}>|^2
\eea
Note that ${\hat\rho}({\hat q})$ is positive, and ${\hat\rho}=0$ when 
${\hat q}$ is not in the light cone. It is also Lorentz
invariant. Thus we write
\bea
{\hat\rho}({\hat q})={\hat\sigma}({\hat q}^2)\theta({\hat q}_0),
~~{\hat\sigma}({\hat q}^2)=0,~~~if~~{\hat q}^2<0
\eea
This is a positive measure. We may separate the expression for the
 VEV of the commutator (\ref{KL}) into two parts:
the single particle state contribution and the rest. Moreover, 
we use the asymptotic state condition to arrive at
\bea
\label{KL1}
<0|[{\hat\phi}(\hat x}),{\hat\phi}({\hat y})]|0>=
i{\hat Z}{\hat\Delta}({\hat x},{\hat y}; m_0)+
i\int_{{\hat m}_1^2}^{\infty}d{\hat m}'^2{\hat{\Delta}
({\hat x},{\hat y}; {\hat m}')
\eea
where ${\hat\Delta}({\hat x},{\hat y};m_0)$ is the VEV of the free field 
commutator, $m_0$ is the mass of the scalar. ${\hat m}_1^2>{\hat M}^2$,
 the multiple
particle threshold.\\
We are in a position to  study several attributes of scattering amplitudes 
in the five dimensional theory such as proving existence of
the Lehmann-Martin ellipse, give a proof of fixed t dispersion relation 
to mention a few. However, these properties
have been derived in a general setting recently \cite{jmjmp1} for 
D-dimensional theories.
The purpose of incorporating the expression for the VEV of the commutator 
of two fields in the 5-dimensional theory
is to provide a prelude to the modification of similar expressions when we 
compactify the theory on $S^1$ as we shall
see in the next section.

\bigskip

\noindent{\bf 3.2. The Compactification of Scalar Field Theory: } 
${\bf R^{4,1} \rightarrow  R^{3,1}\otimes S^1}$

\bigskip

\noindent  In this subsection,  $S^1$ compactification of a spatial 
coordinate of the five dimensional theory is considered. To start with, 
decompose the
five dimensional spacetime coordinates, ${\hat x}^{\hat\mu}$,  as follows:
\bea
 {\hat x}^{\hat\mu}=(x^{\mu}, y)
  \eea
  where $x^{\mu}$ are the four dimensional Minkowski  space coordinates; 
$y$ is the compact coordinate on $S^1$ with periodicity
   $y+2\pi R = y$, $R$ being the radius of $S^1$. We summarize below  
the attributes of this $S^1$
  compactification. The neutral scalar field of mass $m_0$ defined  in $D=5$ 
 manifold  is  now described in the geometry $R^{3,1}\otimes S^1$. 
  We focus on the  free field version {\it such as the  'in' and 'ou't field}, 
  ${\hat\phi}^{in, out}({\hat x})$. The equation of motion is
  $[{\Box}_5+m_0^2]{\hat\phi}^{in,out}({\hat x})=0$. We expand the field 
  \bea
  \label{kk1}
  {\hat\phi}^{in,out}({\hat x})={\hat\phi}^{in,out}(x,y)=\phi^{in,out}_0(x)+
\sum_{n=-\infty, n\ne 0}^{+n=\infty}\phi^{in,out}_n(x)e^{{{in y}\over{ R}}}
  \eea
  Note that $\phi^{in,out}_0(x)$, the so called zero mode, has no 
$y$-dependence. The terms in rest of the series  (\ref{kk1})
  satisfy periodicity in $y$. The  five dimensional Laplacian, ${\Box}_5$, 
is decomposed  as sum two operators: 
    $\Box_4$ and  ${{\partial}\over{\partial y^2}}$ . 
The equation of motion is
  \bea
  \label{kk2}
  [\Box_4 - {{\partial}\over{\partial y^2}}+m_n^2]\phi^{in,out}_n(x,y)=0
  \eea
  where $\phi^{in,out}_n(x,y)=\phi_n^{in,out}e^{{{in y}\over{ R}}}$ 
and $n=0$ term has no $y$-dependence being $\phi_0(x)$; from now on
  $\Box_4=\Box$.
  Here $m_n^2=m_o^2+{{n^2}\over{R^2}}$. Thus we have tower of massive states.
 The momentum associated
  in the $y$-direction is $q_n=n/R$ and is quantized  in the units of 
$1/R$ and it is an additive conserved quantum
  number. We term it as Kaluza-Klein (KK) charge although there is no 
gravitational interaction in the five dimensional theory;  we
   still call it KK reduction.  For the interacting field 
${\hat\phi}({\hat x})$, we can adopt a similar mode expansion. 
   \bea
   \label{kk2x}
   {\hat\phi}({\hat x})={\hat\phi}(x,y)=
\phi_0(x)+\sum_{n=-\infty, n\ne 0}^{n=+\infty}\phi_n(x)e^{{{iny}\over R}}
   \eea
   The equation of motion for the interacting fields is endowed with a 
source term. Thus source current would be expanded
   as is the expansion (\ref{kk2x}). Each field $\phi_n(x)$ will 
have a current, $J_n(x)$ associated with it and source
   current will be expanded as
   \bea
   \label{kk2a}
   {\hat j}(x,y) =j_0(x)+\sum_{n=-\infty, n\ne 0}^{n=+\infty}J_n(x)e^{i{ny/R}}
   \eea
   Note that the set of currents,  $\{J_n(x)  \}$, are the source 
currents associated with the tower of interacting fields
   $\{ \phi_n(x) \}$. These fields carry the discrete KK charge, $n$. 
Therefore, $J_n(x)$ also carries the
   same KK charge. We should keep this aspect in mind when we consider 
matrix element of such currents between
   stated. In future, we might not explicitly display the charge of the 
current; however, it becomes quite obvious
   in the context.\\ 
  The zero mode, $\phi^{in,out}_0$, create their  Fock spaces. 
Similarly, each of the fields $\phi^{in,out}_n(x)$ create  their Fock 
spaces as well.
  For example a state with spatial momentum, ${\bf p}$, energy, $p_0$ and 
discrete momentum $q_n$ (in $y$-direction) is created
  by
  \bea
   \label{kk3}
   A^{\dagger}({\bf p},q_n)|0>=|p,q_n>,~~p_0>0 
   \eea
  {\it Ramark:} The five dimensional theory has a neutral, massive scalar field.
  After the $S^1$ compactification to  the $R^{3,1}\otimes S^1$, 
the spectrum of the resulting theory consists of 
  a massive field of mass $m_0$, associated with the zero mode and tower of
 Kaluza-Klein (KK) states  characterized by a mass and a 'charge',
 $(m_n, q_n)$,
  respectively. We now  discuss the structure of the Hilbert space of
 the compactified theory.\\ 
   {\it The Decomposition of the Hilbert space ${\hat{\cal H}}$:}
 The Hilbert space associated with the five dimensional theory is 
${\hat {\cal H}}$. It is now decomposed as a direct sum of Hilbert
 spaces where each one is characterized by its quantum number $q_n$   
  \bea
  \label{kk4}
  {\hat{\cal H}}=\sum \oplus {\cal H}_n
  \eea
  Thus ${\cal H}_0$ is the Hilbert space constructed from 
$\phi_0^{in,out}$ with charge $q_{n=0}$. This space is built by the
 actions of the creation operators 
  $\{ a^{\dagger}({\bf k}) \}$  acting
  on the vacuum and these states span ${\cal  H}_0$. 
   A single particle state is  $a^{\dagger}({\bf k})|0>=|{\bf k}>$ 
and multiparticle states
  are created using the procedure out lines in (\ref{fock1}) and
 (\ref{fock2}). We can create Fock spaces by the actions of fields
   $\phi_n(x,y)$ with  charge $q_n$ on the vacuum. This space is constructed 
through the action of creation operators
   $\{A^{\dagger}({\bf p}, q_n)\}$. Now  two  state vectors with 
different 'charges'
  are orthogonal to one another 
  \bea
  \label{kk5}
  <{\bf p}, q_{n'}|{\bf p}',q_{n'}>=\delta^3({\bf p}-{\bf p'})\delta_{n,n'}
   \eea
  {\it Remark: } We assume that there are no bound states in the theory 
and all particles
  are stable as mentioned. There exists a possibility that a particle 
with charge $2n$ and mass $m^2_{2n}=m_0^2+{{4n^2}\over{R^2}}$ could be a
  bound state of two particles of charge $n$ and masses $m_n$ each under
 certain circumstances. We have excluded such possibilities
  from the present investigation. \\
  The LSZ formalism can be adopted for the compactified theory. If we keep 
in mind the steps
  introduced above, it is possible to envisage field operators 
$\phi^{in}_n(x)$ and $\phi^{out}_n(x)$ for each of the fields
 for a given $n$.
  Therefore, each Hilbert space, ${\cal H}_n$ will be spanned by the state
 vectors created by operators $a^{\dagger}({\bf k})$, for $n=0$ and
  $A^{\dagger}({\bf p}, q_n)$, for $n\ne 0$.  Moreover, we are in a
 position to define corresponding set of interacting field $\{\phi_n(x) \}$
 which
  will interpolate  into 'in' and 'out' fields in the asymptotic limits.\\
  {\it Remark}: Note that in (\ref{kk1}) sum over $\{n\}$ runs over
 positive and negative integers. If there is a parity symmetry 
$y\rightarrow -y$
  under which the field is invariant we can reduce the sum to positive 
$n$ only. However, since $q_n$ is an additive discrete quantum number, 
  a state with $q_n>0$ could be designated as a particle and the 
corresponding state $q_n<0$ can be interpreted as its antiparticle. Thus
  a two particle state $|p,q_n>|p,-q_n>,~~q_n>0~and ~p_0>0$ is a 
particle 
antiparticle state, $q_n=0$; in other words the sum of the
total charges of the two states is zero.  Thus it has the quantum
number of the vacuum. For example,
  it could be two particle state of $\phi_0$ satisfying energy 
momentum conservation, especially if they appear as intermediate states. 
   \\
   Now return to  the K\"allen-Lehmann representation (\ref{KL}) 
in the present context and utilize the expansion (\ref{kk2x}) in the 
   expression for the VEV of the commutator of two fields defined 
in $D=5$:  $<0|[{\hat\phi}({\hat x}),{\hat\phi}({\hat x}')]|0>$
   \bea
   \label{kk6}
  <0|[{\hat\phi}(x,y),{\hat\phi}(x',y')]|0>=<0|[\phi_0(x)
+\sum_{-\infty}^{+\infty}\phi_n(x,y),~ \phi_0(x')
+\sum_{-\infty}^{+\infty}\phi_l(x',y')]|0> 
   \eea
   The VEV of a commutator of two fields  given by the  spectral 
representation (\ref{KL}) will be decomposed into sum of several commutators
   whose VEV will appear:
   \bea
   \label{kk7}
   <0|[\phi_0(x), \phi_0(x')]|0>,~~<0|[\phi_n(x),\phi_{-n}(x')]|0>,...
   \eea
   Since the vacuum carries zero KK charge, $q_{vac}=0=q_0$, the commutator of 
two fields (with $n\ne 0$) should give rise to zero-charge and 
   only $\phi_n$ and $\phi_{-n}$ commutators will appear.
    Moreover, commutator of fields with different $q_n$ vanish since
 the operators act on states of different Hilbert spaces. Thus
   we already note the consequences of compactification.
 When we wish to evaluate the VEV and insert complete set of intermediate
   states in the product of two operators after opening up the commutators, we note that all states of the entire KK tower can appear
   as intermediate states as long as they respect all conservation laws.
 This will be an important feature in all our computations in
   what follows.

   \bigskip
   
   \noindent{\bf{3.3.  Definitions and Kinematical Variables }}
   
   \bigskip
   
   \noindent  The purpose of this investigation is to derive analyticity
 property of the fixed-$t$ dispersion relations for scattering
   of the KK states carrying nonzero charge i.e. scattering in the 
$q_n\neq 0$ sector. However, we mention in passing the other
   possible processes. These are (i) scattering of states with $q_n=0$ 
states, i.e. scattering of zero modes. (ii)
   The scattering of a state carrying charge $q_n=0$ with a state with 
non-zero KK charge. We have studied 
   reactions (i) and (ii) in I and therefore, we do not wish to 
dwell upon them.\\ 
   We shall define the kinematical variables below. The states carrying 
$q_n\ne 0$ are denoted by $\chi_n$ (from now on a state
   carrying charge is defined with a subscript $n$ and momenta carried 
by external particles are denoted as $p_a, p_b,...$. Moreover, we shall 
consider elastic scattering of states carrying equal charge;
   the elastic scattering of unequal charge particles is just elastic 
scattering of unequal mass states due to mass-charge relationship
   for the KK states.\\   
Let us consider a generic 4-body reaction (all states carry non-zero $n$)
\bea
\label{kk6}
 a + b\rightarrow c + d 
\eea
The particles $( a,  b,  c,  d) $ (the corresponding fields being 
$\chi_a, \chi_b, \chi_c, \chi_d$)
respectively carrying momenta 
$ {\tilde p}_a, {\tilde p}_b, {\tilde p}_c, {\tilde p}_d$; 
these particles may
correspond to the KK zero modes (with KK momentum $q=0$) or particles 
might carry nonzero KK charge. We shall consider
only elastic scatterings. The Lorentz invariant Mandelstam variables are
\bea
\label{kk7}
s=( p_a+ p_b)^2=( p_c+ p_d)^2,~t=( p_a- p_d)^2=( p_b- p_c)^2,~
u=( p_a-p_c)^2=( p_b- p_d)^2
\eea
and $\sum  p^2_a+ p^2_b+ p^2_c+ p^2_d=m_a^2+m_b^2+m_c^2+m_d^2$.
The independent identities of the four particles will facilitate the 
computation of the amplitude so that to keep track
of the fields reduced using LSZ procedure.  
  We list below some relevant (kinematic) variables which will be required in 
future
\bea
\label{kk8}
{\bf M}_a^2,~~{\bf M}_b^2,~~{\bf M}_c^2,~~{\bf M}_d^2
\eea
These correspond to lowest mass two or more particle states which carry 
the same quantum number as that of
particle $a$, $b$, $c$ and $d$ respectively. We  define below six more 
variables 
\bea
\label{kk9}
({\bf M}_{ab}, {\bf M}_{cd}),~~({\bf M}_{ac}, {\bf M}_{bd}),~~
({\bf M}_{ad}, {\bf M}_{bc})
\eea
The variable ${\bf M}_{ab}$ carries the same quantum number as $(a ~and~ b)$
 and it corresponds to two or more particle
states. Similar definition holds for the other five variables introduced above. 
 We define two types of thresholds: (i) the physical threshold, $s_{phys}$,
 and $s_{thr}$. In absence of anomalous thresholds (and equal mass scattering)
 $s_{thr}=s_{phys}$. Similarly, we may define $u_{phys}$ and $u_{thr}$ which 
will be useful when we discuss dispersion relations.  We assume from 
 now on that $s_{thr}=s_{phys}$ and $u_{thr}=u_{phys}$.
Now we outline the derivation of the expression a four point function 
 in the  LSZ formalism.  We  start with 
$| p_d, p_c~out>$ and $| p_b, p_a~in>$ and considers the matrix element 
$< p_d, p_c~out|  p_b, p_a~in>$.
Next we subtract out the matrix element 
$< p_d, p_c~in| p_b, p_a~in>$ to define the S-matrix element.
\bea
\label{kk10}
< p_d, p_d~out| p_b,p_a~in>=&&\delta^3({\bf p}_d-{\bf p}_b)
\delta^3({\bf p}_c-{\bf p}_a)
-{{i}\over{(2\pi)^3}} \int d^4x\int d^4x' \nonumber\\&&
e^{-i( p_a.x- p_cx')}K_xK_{x'}< p_d~out|R(x',x)| p_b~in>
\eea
where $K_x$ and $K_{x'}$ are the four dimensional Klein-Gordon operators and
\bea
\label{kk11}
R(x,x')=-i\theta (x_0-x_o')[\chi_a(x),\chi_c(x')]
\eea
We have reduced fields associated with $a$ and $c$ in (\ref{kk10}). 
In the next step we may reduce all the four fields and in such a reduction
 we shall get VEV of the R-product of four fields which will be operated
upon by four K-G operators. However, the latter form of  LSZ reduction 
(when all fields are reduced) 
is not very useful when we want to investigate the analyticity property 
of the
amplitude in the present context.
 In particular our intent is  to write the  dispersion relation.
 Thus we abandon the idea of reducing all the four fields.\\
{\it Remark: } Note that on the right hand side of the equation 
(\ref{kk10}) the operators act on
 $R\chi_a(x)\chi(x')_c$ and there is a $\theta$-function
in the definition of the R-product. Consequently, the action of 
$K_xK_{x'}$ on $R\chi_a(x)\chi_c(x')$ will produce a term 
$R J_a(x) J_c(x')$. In addition the operation of the two K-G operators 
will give rise to $\delta$-functions and derivatives of $\delta$-functions
and some equal time commutators i.e. there will terms whose coefficients 
are $\delta (x_0-x_0')$. When we consider Fourier transforms of 
the derivatives of these $\delta$-function derivative terms they will be 
transformed to momentum variables. However, the amplitude is a function of
Lorentz invariant quantities. Thus one will get only finite polynomials of 
such variables, as has been argued by Symanzik \cite{kurt}.
His arguments is that  in a local quantum field
theory only finite number of derivatives of $\delta$-functions can appear.
 Moreover, in addition, there are some 
equal time commutators and many of them vanish when we
invoke locality arguments. Therefore, we shall use the relation
\bea
\label{kk12}
K_xK_{x'}R\chi(x)\chi_c(x')=R J_a(x) J_c(x')
\eea
 keeping in mind that there are derivatives of $\delta$-functions and 
some equal time commutation relations which might be present.
 Moreover, since the derivative terms give rise to polynomials in Lorentz
 invariant variables, the analyticity properties of the amplitude
 are not affected due to the presence of such terms. This will be
 understood whenever we write an equation like (\ref{kk12}). 

  \bigskip
  
  \noindent {\bf 4.  Nonforward Elastic Scatting  of $n \ne 0$ 
Kaluza-Klein States}
  
  \bigskip
  
  \noindent  We envisage elastic scattering of two equal mass, 
$m_n^2=m_0^2+{{n^2}\over{R^2}}$,  hence equal charge KK particles 
  and we take $n$ positive. 
  Our first step is to define the scattering amplitude for this reaction 
(see \ref{kk10}) 
\bea
 \label{nn1}
 <p_d,p_c~out|p_b,p_a~in>=&& 4p^0_ap^0_b
\delta^3({\bf p}_d-{\bf p}_b)\delta^3({\bf p}_a-{\bf p}_c) -\nonumber\\&& 
{{i}\over{(2\pi)^3}}
 \int d^4x\int d^4x'e^{-i(p_a.x-p_c.x')}\times \nonumber\\&& 
{\tilde K}_x{\tilde K}_{x'}<p_d~out|{\bar R}(x';x)|p_b~in>
 \eea
  where
\bea
 \label{nn2}
 {\bar R}(x';x)=-i\theta (x_0-x_0')[\chi_a(x),\chi_c(x')]
 \eea  
  and ${\tilde K}_x=(\Box+m_n^2)$. We let the two KG operators act on 
${\bar R}(x;x')$  in the VEV and resulting
  equation is
  \bea
  \label{nn3}
 <p_d,p_c~out|p_b,p_a~in>=&&<p_d,p_c~in|p_b,p_a~in> 
  -{{1}\over{(2\pi)^3}}
 \int d^4x\int d^4x'e^{-i(p_a.x-p_c.x')} \times \nonumber\\&&
  <p_d|\theta(x_0'-x_0)[J_c(x'),J_a(x)]|p_b>
  \eea
  Here $J_a(x)$ and $J_c(x')$ are the source currents associated with 
the fields $\chi_a(x)$ and $\chi_b(x')$ respectively.
  We arrive at (\ref{nn3}) from (\ref{nn1}) with the understanding that 
the $r.h.s.$ of (\ref{nn3}) contains additional
  terms; however, these terms do not affect the study of the analyticity 
properties of the amplitude as alluded to earlier.
We shall define three distributions which are matrix elements of 
the product of current. The importance of these
functions will be evident in sequel   
\bea
 \label{nn4}
    F_R(q)=\int_{\infty}^{+\infty}d^4ze^{iq.z}
\theta(z_0)<Q_f|[J_a(z/2),J_c(-z/2)]|Q_i>
    \eea
 \bea
    \label{nn5}
 F_A(q)=-\int_{\infty}^{+\infty}d^4ze^{iq.z}
\theta(-z_0)<Q_f|J_a(z/2),J_c(-z/2)]|Q_i>
 \eea
 and
 \bea
 \label{nn6}
 F_C(q)= \int_{-\infty}^{+\infty}d^4ze^{iq.z}<Q_f|[J_a(z/2),J_c(-z/2)]|Q_i>
 \eea
Moreover,
\bea
\label{nn7}
F_C(q)=F_R(q)-FA(q)
\eea 
  $|Q_i>$ and $|Q_f>$ are states which carry four momenta and 
these momenta are held fixed. At this stage we treat them as
  parameter;  
 it is elaborated  in ensuing discussions. Let us focus attention on 
the matrix element of the causal commutator defined in (\ref{nn6}). We open
 up the commutator of the currents and introduce a complete set of physical states. Let us assign KK charge $n$ to each of
 the states. Thus the conservation of KK charge only permits those intermediate states which respect the charge
 conservation laws.  The physical complete sets are:
   $\sum_n|{\cal P}_n{\tilde\alpha}_n><{\cal P}_n{\tilde\alpha}_n|={\bf 1}$ 
and
 $\sum_{n'}|{\bar{\cal P}}_{n'}{\tilde\beta}_{n'}><{\bar{\cal P}}_{n'}
{\tilde\beta}_{n'}|={\bf 1}$.
Here $\{{\tilde\alpha}_n, {\tilde\beta}_{n'} \}$
stand for quantum numbers that are permitted for the intermediate states. 
The matrix element defining $F_C(q)$, (\ref{nn7}), assumes the following form
 \bea
\label{nn8}
&&\int d^4ze^{iq.z}\bigg[\sum_n\bigg(\int d^4{\cal P}_n<Q_f|J_a({z\over 2})|
{\cal P}_n{\tilde\alpha}_n>
<{\cal P}_n{\tilde\alpha}_n|J_c(-{z\over 2})|Q_i>\bigg)\nonumber\\&& -
\sum_{n'}\bigg(\int d^4{\bar{\cal P}}_{n'}
<Q_f|J_c(-{z\over 2})|{\bar{\cal P}}_{n'}{\tilde\beta}_{n'}>
<{\bar{\cal P}}_{n'}{\tilde\beta}_{n'}|J_a({z\over 2})|Q_i>\bigg) \bigg]
\eea
We proceed as follows at this point. Let us use translation operations 
judiciously so that the currents do not carry
any dependence in the $z$-variables in their arguments.
  Subsequently, we integrate over 
$d^4z$ which leads to $\delta$-functions. 
\newpage
The expression for $F_C(q)$  now takes the form
 \bea
\label{nn9}
&& F_C(q)=\sum_n\bigg(<Q_f|j_a(0)|{\cal P}_n={{(Q_i+Q_f)}\over 2}-q,
{\tilde\alpha}_n>\times \nonumber\\&& 
<{\tilde\alpha}_n,{\cal P}_n=
{{(Q_i+Q_f)}\over 2}-q|j_c(0)|Q_i>\bigg) \nonumber\\&&
-\sum_{n'}\bigg(<Q_f|j_c(0)|{\bar{\cal  P}}_{n'}=
{{(Q_i+Q_f)}\over 2}+q,{\tilde\beta}_{n'}>\times \nonumber\\&& 
<{\tilde\beta}_{n'},{\bar{\cal P}}_{n'}={{(Q_i+Q_f)}\over 2}+
q|j_a(0)|Q_i>\bigg)
\eea
A few explanatory comments are in order: The momentum of the intermediate 
state ${\cal P}_n$ appearing
in  first term in (\ref{nn9}) is constrained to ${\cal P}_n=({{Q_i+Q_f}\over 2})-q$ after the $d^4z$ integration. Similarly,
${\cal P}_{n'}=({{Q_i+Q_f}\over 2})+q $ in the second term of (\ref{nn9}). The second point is that, in the derivation of
the spectral representation line (\ref{nn9}) for a theory with single scalar field, the physical intermediate states
correspond to the multiparticle states consistent with energy momentum conservation (physical states). For
the case at hand, the intermediate states consist of the entire KK tower as long as these states satisfy
energy momentum conservation constraints and the KK charge conservation rules. We shall discuss the consequences
of this aspect in the sequel.   \\
Let us define
 \bea
\label{nn10}
2A_s(q)=&&\sum_{n'}\bigg(<Q_f|j(0)_a|{\bar{\cal P}}_{n'}={{(Q_i+Q_f)}\over 2}
+q,{\tilde\beta}_{n'}>
\times \nonumber\\&&
<{\tilde\beta}_n',{\bar{\tilde P}}_n={{(Q_i+Q_f)}\over 2}+q|j_c(0)|Q_i>\bigg)
\eea
and 
\bea
\label{nn11}
2A_u=&&\sum_n\bigg(<Q_f|j_c(0)|{\cal P}_n={{(Q_i+Q_f)}\over 2}-q,
{\tilde\alpha}_n>\times
\nonumber\\&&
<{\tilde\alpha}_n,{\cal P}_n={{(Q_i+Q_f)}\over 2}-q|j_l(0)|Q_i>\bigg)
\eea
 {\it Consequences of microcausality:} The Fourier transform of 
$F_C(q), {\bar{F}}_C(z)$, vanishes outside the light cone.
We recall that, 
\bea
\label{nn12}
F_C(q)= {{1}\over 2}(A_u(q)-A_s(q))
\eea
Moreover, $F_C(q)$ will also vanish as function of $q$ wherever, 
both $A_s(q)$ and $A_u(q)$ vanish simultaneously.
We recall that the the intermediate states are physical states and 
their four momenta lie in  the
forward light cone, $V^+$, as a consequence
\bea
\label{nn13}
({{Q_i+Q_f}\over 2}+q)^2\ge 0,~~~({{Q_i+Q_f}\over 2})_0+q_0\ge 0
\eea
and
\bea
\label{nn14}
({{Q_i+Q_f}\over 2}-q)^2\ge 0,~~~({{Q_i+Q_f}\over 2})_0-q_0\ge 0
\eea
The above two conditions, for nonvanishing of $A_u(q)$ and $A_s(q)$ 
implies existence of minimum mass parameters\\ 
(i) $ ({{Q_i+Q_f}\over 2}+q)^2\ge {{\cal M}_+}^2$ 
and (ii) $({{Q_i+Q_f}\over 2}-q)^2\ge {{\cal M}_-}^2 $.\\
 The matrix elements for
$A_s(q)$ and $A_u(q)$ will not vanish and if the two conditions stated above,
pertinent to each of them, are fulfilled.\\
We would like to draw the attentions of the reader to the following 
facts in the context a theory with compactified spatial dimension.
In the case where there is only one scalar field, the sum over 
intermediate physical states as given in (\ref{nn10}) and (\ref{nn11}) 
is the
multiparticles states  permitted by energy momentum conservations. 
However, in the present situation, the contributions to 
the intermediate states are those which come from the KK towers as allowed 
by the charge conservation rules (depending on what
charges we assign to $|Q_i>$ and $Q_f>$ for the elastic scattering) and 
energy momentum conservation. For example, if the initial
states have change $n=1$, then the tower of multiple particle intermediate
 states should have one unit of KK charge. Thus the question
is whether the infinity tower of KK states would contribute? It looks like 
that at the present stage, when we are in the 'linear programme"
framework of the general field theoretic formalism, this issue cannot
 be resolved. As we shall discuss subsequently, when unitarity
constraint is invoked there are only contributions from finite 
number of terms as long as $s$ is finite but can be taken to be very large.\\
In order to derive a fixed-$t$ dispersion relation we have to identify a domain which is free from singularities in the $t$-plane. The first step
is to obtain the Jost-lehmann-Dyson representation for the causal commutator, $F_C(q)$.  We are considering elastic scattering of equal mass particles
i.e. all particles carry same KK charge. Therefore, the technique of Jost and Lehmann \cite{jl} is quite adequate; we do not have to resort to
more elegant and general approach of Dyson \cite{dyson} (see  \cite{jmjmp1} for detail discussions). We shall adhere to notations and discussions of
reference I and present those results in a concise manner.
 As noted in (\ref{nn13}) and (\ref{nn14}), $F_C(q)$ is nonvanishing in those domains. We designate this region
as ${\bar{\bf R}}$,
\bea
\label{nn15}
{\bar{\bf R}}: \bigg\{(Q+q)^2\ge
{{\cal M_+}}^2, Q+q\in V^+ ~ {\rm and} ~ (Q-q)^2 \ge{{\cal M_-}}^2, 
Q-q\in V^+  \bigg\}
\eea
where $Q={{Q_i+Q_f}\over 2}$ and  $V^+$ being the future light cone. 
We need not repeat derivation of the Jost-Lehmann representation here. The
present case differs from the case where only one field is present in the 
following way. Here we are looking for the nearest singularity to
determine the singularity free region. For the case at hand, 
the presence of the towers of KK states is to be envisaged in 
the following perspective. Since we consider equal mass scattering 
the location of nearest singularity will be decided by the lowest
values of ${\cal M}_+$ and ${\cal M}_-$. Let us elaborate this point. 
We recall that there is the tower of KK states appearing as intermediate
states (see (\ref{nn10} ) and (\ref{nn11} )). Thus each new threshold 
could create region of singularity of $F_C(q)$. We are concerned
about the  identification of the singularity free domain. Thus the lowest 
threshold of two particle intermediate state, consistent
with desired constraints, control the determination of this domain 
of analyticity. Therefore,
for the equal mass case, the Jost-Lehmann representation for 
$F_C(q)$ is such that it is nonzero in the region ${\bar{\bf R}}$,
\bea
\label{nn16}
F_C(q)=\int_Sd^4u\int_0^{\infty}d\chi^2\epsilon(q_0-u_0)\delta[(q-u)^2-\chi^2)]
\Phi(u,Q.\chi^2)
\eea
Note that $u$ is also a 4-dimensional vector ({\it not the Mandelstam
variable u}). The domain of integration of $u$ is the region $S$ specified
below
\bea
\label{nn17}
{\bf S}:\bigg\{Q+u \in V^+,~ Q-u \in V^+,~
Max~ [0,{\cal M}_+-\sqrt{(Q+u)^2},{\cal M}_--\sqrt{(Q-u)^2}]\le \chi \bigg\}
\eea
and $ \Phi(u,Q.\chi^2)$ arbitrary.
 Here $\chi^2$ is to be interpreted like a mass parameter. 
Moreover, recall that
the assumptions about the features of the causal function stated above are
the properties we have listed earlier  and $Q$ is already defined above. 
Since the retarded
commutator involves a $\theta$-function, if we use integral representation 
for it (see \cite{jl}) we
derive an expression for the regarded function,
\bea
\label{nn18}
F_R(q)={{i\over {2\pi}}}\int d^4q'\delta^3({\bf q'}-{\bf q})
{{1\over{q_0'-q_0}}}F_C(q'), ~Im~q_0>0
\eea
Moreover, for the retarded function, $F_R(q)$, 
the corresponding Jost-Lehmann representation  reads \cite{jl}
\bea
\label{nn19}
F_R(q)={{i\over{2\pi}}}\int_Sd^4u\int_0^{\infty}d\chi^2
{{\Phi(u,Q,\chi^2)}\over{(q-u)^2-\chi^2}}
\eea
We mention in passing that these integral representations are written 
under the assumption that the functions appearing
inside the integral are such that the integral converges. However, 
if there are polynomial growths asymptotically then
subtraction procedure can be invoked to tame the divergences. It is to 
be borne in mind that these expressions can have
only polynomial behaviors for asymptotic values of the argument as 
we have argued earlier. The polynomial behaviors will not affect the study 
of analyticity
properties. One important observation is that that the singularities 
lie in the complex $q$-plane \footnote{see Itzykson and Zubber 
 \cite{book3} and
Sommer \cite{sommer} for elaborate discussions}. We provide  below 
a short and transparent discussion for the sake of completeness.
The locations of the singularities are found by examining where the 
denominator (\ref{nn19}) vanishes,
\bea
\label{nn20}
(q_0-u_0)^2-(q_1-u_1)^2-(q_2-u_2)^2-(q_3-u_3)^2=\chi^2
 \eea
 We conclude that the the singularities lie on the hyperboloid give by 
(\ref{nn20}) and those points are in domain $\bf S$ as defined in  
(\ref{nn17}). There are points in the hyperboloid which belong to the 
domain $\bf S$. These are called admissible. Moreover, according our
earlier definition,  the domain ${\bar{\bf R}}$ is where $F_C(q)$ is 
nonvanishing (see (\ref{nn15})). Then there  is a domain 
which contains a set of real 
points where $F_C(q)$ vanishes, call it $\bf R$ and this is compliment 
to real elements of ${\bar{\bf R}}$. From the above arguments, we arrive
at the conclusion that $F_C(q)=0$  for every real point belonging to 
${\bf R}$ (the compliment of ${\bar{\bf R}}$). Thus these are the
real points in the $q$-plane where $F_R(q)=F_A(q)$ since $F_C(q)=0$ 
there. Recall the definition of ${\bar{\bf R}}$, (\ref{nn15}). A border
is defined by the upper branch of the parabola given by the 
equation $(Q+q)^2={\cal M_+}^2$ and the other one is given by the equation
for another parabola $(Q-q)^2={\cal M_-}^2$. Now we identify the 
{\it coincidence region} to be the domain bordered by the two parabolae.
It is obvious from the above discussions  that the set 
$\bf S$ is defined by the range of values $u$ and $\chi^2$ 
assume in the admissible
parabola. Now we see that those set of values belong to a subset of 
$(u,\chi^2)$ of all parabolas (recall equation (\ref{nn20}))  
\cite{sommer} and  \cite{jl,dyson} .  In order to transparently 
discuss the location of a singularity, let us go through a few 
short steps as
 the prescription to illustrate essential points.  We discussed about the 
identification of admissible parabola. The amplitude is function of
Lorentz invariant kinematical variables; therefore, it is desirable 
to express the constraints and equations in terms of those variables
eventually. Let us focus on $Q\in V^+$ and choose a Lorentz frame such 
that four vector $Q=(Q_0,{\bf 0})$ where $\bf 0$ stands for the
{\it three} spatial components of $Q$. Next step is to choose four 
vector $q$ appropriately to exhibit the location of singularity in 
a simple way.
This is achieved as follows: choose one spatial component of 
$q$ in order to identify the position of the singularity in this variable and
treat $q_0$ and the rest of the components of $q$ as parameters and 
hold them fixed \cite{sommer}. We remind the reader that all the variables 
appearing in the Jost-Lehmann representation for $F_C(q)$ and 
$F_R(q)$ are Lorentz invariant objects. Thus going to a specific 
frame will not alter
the general attributes of the generalized functions. 
If we solve for $q_1^2$ in (\ref{nn20}) after obtaining an expression 
for $q_1^2$
\bea
\label{nn21}
q_1=u_1\pm i\sqrt{\chi_{min}^2(u)-(q_0-u_0)^2+(q_2-u_2)^2+(q_3-u_3)^2+\rho}, \rho>0
\eea
We remind that the set of points$\{u_0,u_1,u_2,u_3 ; 
\chi_{min}^2= min~\chi^2 \}$ lie in $\bf S$. The above exercise 
has enabled us
to identify the domain where the singularities might lie with 
the choice for the variables $Q$ and $u$ we have made. We are dealing with
the equal mass case and note that the location of the singularities are 
symmetric with respect to the real axis. We now examine a further
simplified scenario where the coincidence region is bounded by two 
branches of hyperboloids so that ${\cal M}_+^2={\cal M}_-^2={\cal M}^2$.
Now the singular points are
\bea
\label{nn22}
q_1=u_1\pm i\sqrt{Min~[\chi_{min}^2-u_0^2+u_2^2+u_3^2]+\rho},\rho>0
\eea
For the case under considerations:  $(Q+q)^2=(Q-q)^2={\cal M}^2$,
and  
\bea
\label{nn23}
 q_1=u_1\pm i\sqrt{({\cal M}-\sqrt{Q^2-u_1^2)^2}+\rho},\rho>0
 \eea 
Now we can utilize this analysis to present a derivation of the 
Lehmann ellipse. The essential difference between the present investigation
in this context with the known results is that now we have to deal with 
several thresholds for identification of the coincidence
regions. These thresholds are the multiparticle states in various channels 
as discussed earlier as introduced in Section 3 through
the two equations (\ref{kk8}) and (\ref{kk9}). Their relevance is already 
reflected in the spectral representations, (\ref{nn10}) and (\ref{nn11}),
when we introduced complete set of intermediate states. 
We remark in passing that the presence of the excited KK states do not
shrink the singularity free regions. Therefore, the domain we have 
obtained is the smallest domain of analyticity; nevertheless, we feel
that in order to arrive at this conclusion the entire issue had to be 
examined with care.

\bigskip

\noindent {\it The Lehmann Ellipses}

\bigskip

\noindent  Our goal is to derive fixed-$t$ dispersion relations. 
We have noted that as $s\rightarrow s_{thr}$,   cos$\theta$ goes out of the
physical region $-1\le cos\theta \le +1$,  $\theta$  being  the $c.m.$ angle
   when  we wish to hold $t$ fixed. We choose the following
kinematical configuration in order to derive the Lehmann ellipse. 
For the case at hand i.e. elastic scattering of equal (nonzero) charge KK
states, hence particles of equal mass. Here  $(a,b)$ and $(c,d)$ are 
respectively the incoming and outgoing particles. They are assigned the
following energies and momenta in the $c.m.$ frame: 
\bea
\label{nn24}
p_a=(E_a,~{\bf k}),~~p_b=(E_b,-{\bf k}),~~p_c =(E_c,~{\bf k}'),~~p_d=(E_d,-{\bf k}')
\eea    
$\bf k$ is the $c.m.$ momentum, $|{\bf k}|=|{\bf k}'|$, 
$E_a=\sqrt{(m_a^2+{\bf k}^2)}$, $E_b=\sqrt{(m_c^2+{\bf k}^2)}$, 
$E_c=\sqrt{(m_c^2+{\bf k}'^2)}$ and
$E_d=\sqrt{(m_d^2+{\bf k}'^2)}$. Although all the particles, 
$(a,b,c,d)$,  are identical,  we keep labeling them as individual one for 
the purpose which
will be clear shortly. Thus $E_a=E_b$ and $E_c=E_d$ and 
${\hat{\bf k}}.{\hat{\bf k}}'=cos\theta$.  It is convenient to choose the
following coordinate frame for the ensuing discussions.
\bea
\label{nn25}
p_a=({\sqrt s},+{\bf k},~{\bf 0}),~~p_b=({\sqrt{s}},-{\bf k},~{\bf 0})
\eea
$\bf 0$ is the two spatial components of vector $\bf k$ and
\bea
\label{nn26}
p_c=({\sqrt{s}}, +kcos\theta,+ksin\theta,0)~~p_d=
({\sqrt s},-kcos\theta, -ksin\theta,0)
\eea
with $k=|{\bf k}|=|{\bf k}'|$.  Thus, $s=(p_a+p_b)^2=(p_c+p_d)^2$
\bea
\label{nn27}
q={1\over 2}(p_d-p_c)=(0,-kcos\theta,-ksin\theta,0),~~
P={1\over 2}(p_a+p_b)=({\sqrt s},0,0,0)
\eea
With these definitions of $q$ and $P$, when we examine the conditions 
for nonvanishing of the spectral representations of  $A_s$ and $A_u$ 
we arrive  at 
\bea
\label{nn28}
(P+q)^2>{{\cal M}_+}^2,~{\rm for}~A_s\ne 0,~~(P-q)^2>{{\cal M_-}^2},~{\rm for}~A_u\ne 0
\eea
Thus the coincidence region is given by the condition
\bea
\label{nn29}
(P+q)^2<{{\cal M}_+}^2,~~(P-q)^2<{{\cal M}_-}^2
\eea
We are dealing with the equal mass case; therefore, 
${{\cal M_+}}^2={{\cal M}_-}^2={\cal M}^2$. We conclude from the energy 
momentum 
conservation constraints  (use the expressions for $P$ and $q$) 
that $p_c^2=(P-q)^2<M_c^2$ and $p_d^2=(P+q)^2<M_d^2$ in this region.
Moreover, $(p_a-p_c)^2=(P-q-p_a)^2<{{\cal M}_{ac}}^2$ and 
$(p_a+p_d)^2=(P-q-p_a)^2<{{\cal M}_{ad}}^2$. We also note that 
$(P-q)\in V^+$ and
$(P+q)\in V^+$. The admissible hyperboloid is 
$(q-u)^2=\chi_{\min}^2+\rho, \rho>0$ with 
$({{p_a+p_b}\over 2}\pm u)\in V^+$.  $\chi_{min}^2$ assumes the 
following form for the equal mass case,
\bea
\label{nn30}
\chi_{min}^2=Max~ \bigg\{0, {\cal M}-\sqrt{({{(p_a+p_b)}\over 2}+u)^2}, 
{\cal M}-\sqrt{({{(p_a+p_b)}\over 2}-u)^2} \bigg\}
\eea
Notice that ${\cal M}$ appearing in the second term of the curly 
in (\ref{nn30}) is the mass of two or more multiparticle states carrying 
the quantum
 numbers of particle $c$; whereas ${\cal M}$ appearing in the third term 
inside the curly bracket is the mass of two or more multiparticle states
 carrying the quantum numbers of particle $d$.
In the present case $\cal M$ has the same quantum number as that of the 
incoming state carrying KK charge $n$. Thus, in this sector, we can 
proceed to show 
the existence of the small Lehmann Ellipse (SLE). It is not necessary 
to present the entire derivation here.  The extremum of the ellipse
is given by
\bea
\label{nn31}
cos\theta_0=\bigg(1+{{(M_c^2-m_c^2)(M_d^2-m_d^2)}\over{k^2(s-M_c^2-M_d^2)}}
 \bigg)^{1/2}
\eea
We note that $M_c=\sqrt{m_n^2+m_0^2}$ is the mass of the lowest multiparticle
 state (one particle with KK charge $one$ and another with KK charge $zero$;
 moreover, $M_c=M_d$. Thus the denominator is $k^2s$. 
 \bea
 \label{nn32}
cos\theta_0=\bigg(1+{{9m_0^4}\over{k^2s}}\bigg)^{1/2}
\eea
It will be a straightforward work to derive the properties of the 
large Lehmann Ellipse (LLE) by reducing all the four fields in 
the expression
for the four point function as is the standard prescription. also note that
 the value of $cos\theta (s)$ depends on $s$.
A natural question to ask is: what is the role of the KK towers?\\

\noindent {\it Important Remark:} The first point to note is that in the 
presence of the other states of KK tower, we have to carry out the same 
analysis as above for
each sector. Notice, however, each multiparticle state composed of KK 
towers has to have the quantum numbers of $c$ (same as $d$ since we
consider elastic channels of equal mass scattering). Thus if $c$ carries 
charge $n$, then a possible KK state could be $q+l+m=n$ since KK 
charges
can be positive and negative. The second point is when we derive the 
value of $cos\theta_0$, for each such case, it is rather easy to work 
out that
value will be away from original expression (\ref{nn31}). 
Thus the nearest singularity in $cos\theta$ plane is given by the expression
(\ref{nn32}) although there will be Lehmann ellipses 
associated with higher KK towers. 
\\
Consequently, when we expand the scattering amplitude in partial waves 
(in the Legendre polynomial basis) the domain of convergence
is to be identified. This domain 
of analyticity is enlarged (earlier it was only physically permitted 
values of $-1\le cos\theta\le +1$) to a region which is an ellipse whose semimajor 
axis is given by (\ref{nn32}). Moreover, the absorptive part of the 
scattering amplitude has a domain of convergence beyond $cos\theta=\pm 1$;
  it converges
inside the large Lehmann ellipse (LLE).
Therefore, we are able to write fixed-$t$ dispersion relations as long as 
$t$ lies in the following domain
\bea
\label{nn33}
|t|+|t+4k^2|<4k^2cos\theta_0
\eea
The absorptive parts
$A_s$ and $A_u$ defined on the right hand and left hand cuts respectively, for
$s'>s_{thr}$ and $u'>u_{thr}$ are holomorphic in the LLE. Thus, assuming no
subtractions 
\bea
\label{disperse}
F(s,t)={1\over \pi}\int _{s_{thr}}^{\infty}{{ds'~A_s(s',t)}\over{s'-s}}
+{1\over \pi}\int _{u_{thr}}^{\infty}{{du'~A_u(u',t)}\over{u'+s}-4m^2+t}
\eea
We shall discuss the issue of subtractions in sequel. We remark in passing 
that crossing has not been proved explicitly in this investigation.
However, it is quite obvious from the preceding developments, it will not 
be hard to prove crossing either from the prescriptions of
Bremmermann, Oehme and Taylor \cite{bot} or from the procedures of 
Bross, Epstein and Glaser \cite{beg1}.

\bigskip

\noindent  { \bf 5. Unitarity and Asymptotic Behavior of the Amplitude} 

\bigskip

\noindent In this section we shall explore the consequences of unitarity 
as mentioned earlier. The investigation so far has followed
what is known as {\it the linear program} in axiomatic field theory. 
All our conclusions about the analyticity properties of the scattering
amplitude are derived from micro causality, Lorentz invariance, 
translational invariance and axioms of LSZ. Note that unitarity of the
$S$-matrix is a nonlinear relationship and it is quite powerful. 
For example, the positivity properties of the partial wave amplitude
follows as a consequence. First we utilize unitarity in a new context in 
view of the fact that there are infinite towers of KK states
in the spectral representation of $F_C(q)$ and the representation for 
$F_R(q)$.  \\
  
  Let us define the $\bf T$-matrix as follows:
  \bea
  \label{nn56}
  {\bf S}={\bf 1}-i{\bf T}
  \eea 
  The unitarity of the S-matrix, 
${\bf {SS^{\dagger}}}={\bf {S^{\dagger}S}}={\bf 1}$ yields
  \bea
  \label{nn57}
  ({\bf{T^{\dagger}}}-{\bf{T}})=i{\bf{T^{\dagger}T}}
  \eea
  In the present context,  we consider the matrix element  for the 
reaction $a+b\rightarrow c+d$. Note that on
  $L.H.S$ of ({\ref{nn5}) it is taken between ${\bf{T^{\dagger}-T}}$. 
We introduce a complete set of physical states
  between ${\bf {T^{\dagger}T}}$.   For the elastic case with all
 particles of KK charge, $n$, the unitarity relation is
  \bea
  \label{nn58}
  <p_d,p_c~in|{\bf{T^{\dagger}}}-{\bf T}|p_b,p_a~in>=
i\sum_{{\cal N}}<p_d,p_c~in|{\bf{T^{\dagger}}}|{\cal N}>
<{\cal N}|{\bf T}|p_b,p_c~in>
  \eea
  The complete of states  stands for 
$|{\cal N}> =|{\cal P}_n{\tilde\alpha}_n>$.  The unitarity relation reads,
  \bea
  \label{nn59}
  T^*(p_a,p_b;p_c,p_d)-T(p_d,p_c;p_b,p_a)=
2\pi i\sum_{\cal N}\delta(p_d+p_c-p_n)T^*(n;p_c,p_d)T(n;p_b,p_a)
  \eea
  We arrive at an expression like the second term of the $R.H.S$ of 
(\ref{nn1}) after reducing two fields. If we reduce a single field
  as the first step (as is worked out in text books) there will be a 
single KG operator acting on the field and eventually
  we obtain matrix element of only a single current.  The $R.H.S.$ of 
(\ref{nn7}) has matrix element like (for example)
  $p_a+p_b\rightarrow p_n$. Thus we can express it as 
\footnote{ We  adopt the arguments and procedures of Gasiorowicz in 
these derivations}
    \cite{gasio}
  \bea
  \label{nn60}
  \delta(p_n-p_a-p_b)T(n:p_b,p_a)= 
(2\pi)^{3/2}<n~out|J_a(0)|p_b>\delta(p_n-p_a-p_b)
  \eea
  After carrying out the computations we arrive at
  \bea
  \label{nn61}
  T(p_d,p_c;p_b,p_a) -T^*(p_d,p_c;p_b,p_a)=&&\sum_{\cal N}
\bigg[\delta(p_d+p_c-p_n)\times \nonumber\\&& 
T(p_d,p_c;n)T^*(n;p_b,p_a)-\nonumber\\&&
  \delta(p_a-p_c-p_n)\times \nonumber\\&& 
T(p_d,-p_c;n)T^*(p_d,-p_c;n) \bigg]
  \eea
  Let consider the scattering amplitude for the reaction under considerations.
  \bea
  \label{nn62}
  F(s,t)=i\int d^4xe^{i(p_a+p_c).{{x}\over 2}}
\theta(x_0)<p_d|[J_a(x/2),J_c(-x'/2)]|p_b>
  \eea
  We evaluate the imaginary part of this amplitude, $F(s,t)$
  \bea
  \label{nn63}
  Im~F(s,t)=&&{{1}\over{2i}}(F-F^*)\nonumber\\&&
  ={{1}\over2}\int d^4xe^{i(p_a+p_c).{{x\over 2}}}
<p_d|[J_a(x/2),J_c(-x/2)]|p_b>
  \eea
  Note that $F^*$ is invariant under interchange $p_b\rightarrow p_d$ and 
also $p_d\rightarrow p_b$; moreover, $\theta(x_0)+\theta(-x_0)=1$.
  We open up the commutator of the two currents in (\ref{nn11}). 
Then introduce a complete set of physical states 
$\sum_{\cal N}|{\cal N}><{\cal N}|=1$. Next we
  implement  translation operations in each of the 
(expanded) matrix elements to  express  arguments of each current as  
$J_a(0)$ and $J_c(0)$
  and finally integrate over $d^4x$  to get the $\delta$-functions. 
As a consequence  (\ref{nn11}) assumes the form
  \bea
  \label{nn65}
  F(p_d,p_c;p_b,p_a)-F^*(p_b,p_a;p_c,p_d)=&&
2\pi i\sum_{\cal N}\bigg[\delta(p_d+p_c-p_n)
F(p_d,p_c;n)F^*(p_a,p_b;n)\nonumber\\&&
  -\delta(p_a-p_c-p_n)\times \nonumber\\&& 
F(p_d,-p_a;n)F^*(p_b,-p_c;n)\bigg]
  \eea
  This is the generalized unitarity relation where all external 
particles are on the mass shell. Notice that the first term on 
the $R.H.S$ of the
  above equation is identical in form to the $R.H.S.$ of (\ref{nn9}); 
the unitarity relation for ${\bf T}$-matrix. The first term in (\ref{nn12}) 
 has
  the following interpretation: the presence of the $\delta$-function 
and total energy momentum conservation implies
  $p_d+p_c=p_n=p_a+p_b$. We identify it as the $s$-channel 
process $p_a+p_b\rightarrow p_c+p_d$.\\
  Let us examine the second term of (\ref{nn12}).  Recall that the 
unitarity holds for the $S$-matrix when all external particles are on shell
  (as is true for the $T$-matrix). The presence of the $\delta$-function 
in the expression ensures that the intermediate physical states will 
contribute for
  \bea
  \label{nn66}
  p_b+(-p_c)=p_n=p_d+(-p_a)
  \eea
  The masses of the intermediate states must satisfy
  \bea
  \label{nn67}
  {\cal M}_n^2=p_n^2=(p_b-p_c)^2
  \eea  
  It becomes physically transparent if we choose the Lorentz frame where  
particle  $'b'$ is at rest i.e. $p_b=(m_b, {\bf 0})$; thus
  \bea
  \label{nn68}
  {\cal M}_n^2=2m_b(m_b-p_c^0),~~p_c^0>0
  \eea
  since $m_b=m_c$ and 
$p_c^0={{\sqrt{m_c^2+{\bf p}_c^2}}}={{\sqrt{m_b^2+{\bf p}_c^2}}} $;  
${\cal M}_n^2<0$ in this case.
  We recall that all particles carry KK charge $n$ and hence the 
mass is $m_b^2=m_n^2=m_0^2+{{n^2}\over{R^2}}$. The intermediate state
  must carry that quantum number. In conclusion, the second term of 
(\ref{nn12}) does not contribute to the $s$-channel reaction. There is
  an important implication of the generalized unitarity equation: 
Let us look at the crossed channel reaction
  \bea
  \label{nn69}
  p_b+(-p_c)\rightarrow p_d+(-p_a);~~ -p_a^0>0,~ and~ -p_c^0>0
  \eea
  Here $p_b$ and $p_c$ are incoming (hence the negative sign for $p_c$) 
and $p_d$ and $p_a$ are outgoing. The second matrix element
  in (\ref{nn12}) contributes to the above process in the configurations 
of the four momenta of these particle;
   whereas the first term in that equation does not if we follow 
the arguments for
  the $s$-channel process. \\
  {\it Remark}: We notice the glimpses of crossing symmetry here.  
Indeed, the starting
  point will be to define $F_C(q)$ and look for the coincidence region.
 Notice that  $q$ is related to physical momenta of external particles when
  $|Q_i>$ and $|Q_f>$ are identified with  the momenta of the
 'unreduced' fields. Indeed, we could proceed to prove crossing  
symmetry for the 
  scattering process; however,  it is not our present goal.\\
   {\it An important observation is  in order}: \\
     We could ask whether entire Kaluza-Klein tower  of states
  would appear as intermediate states in the unitarity equation.
  It is obvious from the unitarity equation (\ref{nn12}) that for the 
$s$-channel process, due to the presence of the energy momentum
  conserving $\delta$-function, $p_n^2={\cal M}_n^2= (p_a+p_b)^2$; 
consequently, not all states of the infinite KK tower will contribute
  to the reaction in this,   ($s$), channel. Therefore the sum would 
terminate after finite number of terms, even for very large $s$ as 
long as it is finite.
   Same argument also holds for the crossed channel process. Thus unitarity 
constraint settles the issue of the contributions of KK towers
   as we alluded to in the previous section in the context of the spectral 
representation of $F_R(q)$, $F_A(q)$ and $F_C(q)$.  
  \\
  Let us turn the attention to the partial wave expansion of the amplitude 
and the power of the positivity property of absorptive part of the
  amplitude. We recall that the scattering amplitude admits a partial 
wave expansion
  \bea
  \label{uni1}
  F(s,t)= {{\sqrt s}\over k} \sum_{l=0}^{\infty}(2l+1)f_l(s)P_l(cos\theta)
  \eea
  where $k=|{\bf k}|$ and $\theta$ is the $c.m.$ scattering angle. 
The expansion converges inside the Lehmann ellipse with with
  focii at $\pm 1$ and semimajor axis $1+{{const}\over {2k^2}}$. 
Unitarity  leads to the positivity constraints on the partial wave
  amplitudes
  \bea
  \label{uni2}
  0\le |f_l(s)|^2\le Im~f_l(s)\le 1
  \eea
  As is well known, the semimajor axis of the Lehmann ellipse shrinks and 
$s$ grows. Recall that derivation of the Lehmann ellipse is
  based on the {\it linear program}.  Martin \cite{martin1}  has proved an 
important theorem. It is known as the procedure for the enlargement of 
the domain of analyticity.
  He demonstrated that the scattering amplitude is analytic in the 
topological product of the domains $D_s\otimes D_t$. This domain is
  defined by $|t|<{\tilde R}$, ${\tilde R}$ being independent of $s$ and 
$s$ is outside the cut $s_{thr}+\lambda=4m_n^2+\lambda, \lambda>0$. In order
  to recognize the importance of this result, we briefly recall the theorem 
of BEG \cite{beg2}. It is essentially  the study of the analyticity property 
of the
  scattering amplitude $F(s,t)$. It was shown that in the neighborhood of 
any point $s_0$, $t_0$  $ -T<t_0\le 0$, $s_0$ outside the cuts, there is
  analyticity in $s$, and $t$ in a region
  \bea
  \label{uni3}
  |s-s_0|<\eta_0(s_0,t_0),~~|t-t_0|<\eta_0(s_0,t_0)
  \eea
  The amplitude is analytic.
  Note the following features of BEG theorem: it identifies the domain of 
analyticity; however, the size of this domain may vary as $s_0$ and $t_0$
  vary. Furthermore, the size of this domain might shrink to zero; in 
other words, as $s\rightarrow 0$, $\eta(s)$ may tend to zero. The importance 
of Martin's
  theorem lies in his proof that $\eta(s)$  is bounded from below 
i.e. $\eta (s)\ge {\tilde R}$ and ${\tilde R}$ is $s$-independent. 
It is unnecessary to repeat
  the proof of Martin's theorem here. Instead, we shall summarize the 
conditions to be satisfied by the amplitude as stated by Martin \cite{martin1}.
  \\
  {\it Statement of Martin's Theorem:} 
   {If following
requirements are satisfied by the elastic amplitude\\
I. $F(s,t)$ satisfies fixed-t dispersion relation in s with finite number of
subtractions ($-T_0\le t\le 0$).\\
II. $ F(s,t)$ is an analytic function of the two Mandelstam variables, 
$s$ and
$t$, in a neighborhood of $\bar s$ in an interval below the threshold,
$4m_n^2-\rho<{\bar s}<4m_n^2$ and also in some neighborhood of $t=0$,
$|t|<R({\bar s})$. This statement hold due to the work of
Bros, Epstein and Glaser \cite{beg1,beg2}.\\
III. Holomorphicity of $A_s(s',t)$ and $A_u(u',t)$: The absorptive parts of
$F(s,t)$ on the right hand and left hand cuts with $s'>4m_n^2$ and $u'>4m_n^2$
are holomorphic in the LLE. \\
IV. The absorptive parts $A_s(s',t)$ and $A_u(u',t)$, for $s'>4m_n^2$ and
 $u'>4m_n^2$ satisfy the following positivity properties
\bea
\label{uni4}
\bigg|{\bigg({\partial\over{\partial t}}}A_s(s',t)\bigg)^n\bigg|
\le {\bigg({\partial\over{\partial t}}\bigg)^n} A_s(s',t)\bigg|_{t=0},~~
-4k^2\le t\le 0
\eea
and
\bea
\label{uni5}
\bigg|{\bigg({\partial\over{\partial t}}}A_u(u',t)\bigg)^n\bigg|
\le {\bigg({\partial\over{\partial t}}\bigg)^n} A_u(u',t)\bigg|_{t=0},~~
-4k^2\le t\le 0
\eea
where ${\bf k}$ is the {\it c.m. } momentum.
Then $F(s,t)$ is analytic in the quasi topological product of the domains 
$D_s\otimes D_t$. (i) $s\in~cut-plane$: $s\ne 4m_n^2+\rho,\rho>0$ and
(ii) $|t|<{\tilde R}$, there exists some ${\tilde R}$ such that 
dispersion relations are valid for $|t|<{\tilde R}$, independent of $s$.
We may follow the standard method to determine ${\tilde R}$. The polynomial 
boundedness, in $s$,  can be asserted by invoking the simple arguments
 presented earlier. Consequently, a dispersion relation can be written 
down for $F(s,t)$ in the domain $D_s\otimes D_t$.  The importance of Martin's
 theorem is appreciated from the fact that it implies that the $\eta$ of 
BEG is bounded from below by an $s$-independent $\tilde R$. Moreover, value
 of $\tilde R$ can be determined by the procedure of Martin 
(see \cite{sommer} for the derivations). \\
 We shall list a few more results as corollary without providing detailed 
computations: \\
 (i) It can be proved that the partial wave expansion can be expressed as 
sum of two terms
 \bea
 \label{uni6}
 F(s,t)= {\cal S}_1+{\cal S}_2
 \eea
 \bea
 \label{uni6}
 {\cal S}_1={{\sqrt s}\over k}\sum_{l=0}^L(2l+1)f_l(s)P_l(1+{{t\over{2k^2}}})
 \eea
 \bea
 \label{uni7}
 {\cal S}_2={{\sqrt s}\over k}\sum_{L+1}^{\infty}
(2l+1)f_(s)P_l(1+{{t\over{2k^2}}})
 \eea
 where $L=const. {\sqrt s}log s$ is the cut off which is derived from 
the convergence of the partial wave expansion inside
 the Lehmann-Martin ellipse and the polynomial boundedness of the amplitude. 
The partial sum ${\cal S}_2$ has subleading 
 contributions to the amplitude compared to ${\cal S}_1$; in fact 
${{\cal S}_1\over{{\cal S}_2}}\rightarrow (log~s)^{-1/4}$ for asymptotic
 $s$ apart from some innocent $t$-dependent prefactor; as is well known. \\
 (ii) {\it The Bound on $\sigma_t$}: The analog of Froissart-Martin bound 
can be \cite{jm2} obtained in that $\sigma_t(s)\le const. (log~s)^2$. 
The constants appearing determining
 $L$ and in derivation of the Froissart-Martin bound can be determined in 
terms of $\tilde R$ and we have refrained from giving those
 details.\\
 (iii){\it Number of subtractions}:  Once we have derived (i) and (ii) it 
is easy to prove the Jin-Martin \cite{jm}  bound which states that the 
amplitude requires at most two subtractions.
 This is achieved by appealing to the existence of fixed-$t$ dispersion 
relation relations and to Phragman-Lindelof theorem.  \\
 We would like to draw the attention of the reader to the fact  that  
a field theory defined on the manifold $R^{3,1}\otimes S^1$ whose spectrum 
consists
 of a massive scalar field and a tower of Kaluza-Klein states satisfies 
nonforward dispersion relations. This statements begs certain clarifications.
 The theory satisfies LSZ axioms. The analyticity properties can be derived 
in the {\it linear program} of axiomatic field theory which
 leads to the proof of the existence of the Lehmann ellipses. The role of the KK tower is to be assessed in this program. Once we invoke
 unitarity constraint stronger results follow and the enlargement of 
the domain of analyticity in $s$ and $t$ variables can be established.

\bigskip

\noindent {\bf 6. Proposal to Explore Decompactification of Extra 
Dimensions} 

\bigskip

\noindent 
In this section we would like to examine a possibility of exploring
the signature of large extra dimension in high energy
collisions of hadrons at LHC. There has been intense phenomenological
activities to look for evidence of extra
spatial dimensions. If the radius of extra dimension is large then
excited states would be produced in proton-proton
scatterings at LHC. There has been proliferation of model suggesting detection
of these exotic particles. There is no
conclusive experimental evidence so far to confirm that there are
compact spatial dimensions with large radius
of compactification.  We refer to the papers cited in the introduction
section \cite{anto,luest}.\\
It was suggested sometime ago \cite{jmjmp15} that precision measurements of
total cross sections at high energy
might be another way to explore whether extra compact dimensions are
 decompactified at LHC energy and beyond.
The proposal is based on the following idea. It is well known that
the total cross section, $\sigma_t$, should respect
the Froissart-Martin bound i.e. it cannot grow faster
than $log^2 s$ at asymptotic energies. If this bound is violated
then axioms of axiomatic local quantum field theory would face
serious problems. There is rigorous derivation
of a bound on total cross section in field theories which live
in higher dimensions \cite{jmjmp}, $D>4$. A hermitian massive
field theory was considered in D-dimensions. The axiomatic LSZ technique
was adopted to investigate the analyticity
properties of the four point amplitude \cite{jmjmp} as has been alluded
to in the introduction section. A bound was derived
\bea
\label{sec5.1}
\sigma_t\le {\tilde C}log^{D-2}{s\over{s_0}}
\eea
where D is the number of spacetime dimensions and $\tilde C$ is a constant,
determined from the first principles.
Note that, for $D=4$ we recover the Froissart-Martin bound.
 Now consider the following situation. Suppose an extra
compact dimension decompactifies at the LHC energy regime.
 Then the energy dependence of total cross section
is not necessarily bounded by $log^2(s)$ and one
would conclude that Froissart-Martin
bound is violated. However,  for
a five dimensional flat Minkowski spactime, the bound on total
cross section is $\sigma_t^{D=5}\le {\tilde C}log^3(s)$.
This bound is derived from LSZ axioms. In such a situation,
 should energy dependence of $\sigma_t$ exhibit
a behavior violating the Froissart-Martin we should refrain
from challenging axioms of local field theory.
The reason is that the Froissart-bound-violating-behavior of $\sigma_t$
 might have a different origin.\\
Nayak and Maharana \cite{nm} have examined this issue recently.
 The first point to note is that the fit to
high energy cross sections has been presented in the particle data group
(PDG) data book \cite{pdg}. They fit the
data which respects Froissart-Martin bound and most of the analysis
 also fit the data with a term of the form
$log^2s$ with a constant coefficient.
We shall discuss this aspect later in this section.
There is an analysis \cite{menon}  which fitted
the data from laboratory energy of 5 GeV to LHC energy and included the cosmic
ray data as well. They claim to have fitted the data with a
Froissart-bound-violating energy dependent term.  It is worth
mentioning that the number of data points in the 'low energy range'
  (i.e. 5 GeV to below ISR energy) are vast compared
to the data points from ISR range to LHC. Moreover, those 'low energy region'
are measured with better precisions.
Therefore, when one adopts a fitting formula and goes for the
$\chi^2$ minimization program these data points
primarily control the minimization procedure. Furthermore,
the fitting procedure and other techniques adopted by them \cite{menon}
have been subject to criticism by Block and Halzen \cite{bh}. We have no
remarks to offer on this issue.\\
We have adopted a different strategy to test whether
Froissart-Martin bound is violated in high energy scattering.
We argue that it is best to test the validity of the aforementioned
bound starting from an energy domain where
$\sigma_t(s)$ start rising with energy. Then one can go all the way up
to the cosmic energy domain.
Our proposal is to consider a set of data from an energy range where
the total cross section starts
growing with energy up to LHC energy and beyond. If we focus only
on $\sigma_t^{pp}$ then the number of data points are
quite limited. We include $\sigma_t^{p\bar p}$ data from ISR,
CERN SPS collider,
Tevatron and fit the total set of data points
which is quite substantial. We justify inclusion of the
$\sigma_t^{p\bar p}$ along with $\sigma_t^{pp}$ data on the following ground.
We invoke the Pomranchuk theorem \cite{pom}. The theorem, in its original form, stated that particle-particle and particle-antiparticle total
cross sections tend to equal values at asymptotic energies.
We recall, Pomeranchuk assumed that the two cross sections attain
constant values at high energy. The total cross sections for $pp$ and
$p\bar p$ started rising from ISR energies. Therefore,
the Pomeranchuk theorem had to be reexamined. The bound on total cross section, $\sigma_t\le log^2s$  has been
proved from analyticity and unitarity of S-matrix in the axiomatic
field theories. It requires additional reasonable assumptions \cite{andre85}
 to derive
behavior of $\Delta\sigma=\sigma_t^{pp}-\sigma_t^{p\bar p}$ for
asymptotic s and show under which circumstances $
\Delta\sigma\rightarrow 0$ as $s$ approaches the limit
$s\rightarrow \infty$. The test of Pomeranchuk theorem
comes from ISR experiments since it measures $pp$ and $p\bar p$ total
cross sections in the same energy domain. Note that  the SPS
($p\bar p $)collider
at CERN and the Tevatron at FERMILAB measure $\sigma_t^{p\bar p}$.
The LHC measures $\sigma_t^{pp}$. It is noteworthy  that
$\Delta\sigma$ shows the tendency to decrease with energy in the energy
regime covered by ISR. Therefore, we feel that  it is quite justified
to combine the high energy total cross sections of $pp$ and $p\bar p$ and
fit the total cross sections. \\
Now we discuss our fits to total cross section \cite{nm}.
We choose the following parameterization to fit the combined data.
\bea
\label{ourfit}
\sigma_t=H log^{\alpha}\left({s\over{s_o}}\right)+P
\eea

$H$ and $P$ are the Heisenberg and Pomeranchuk constants, respectively.
P is the contribution of the Pomeranchuk
trajectory in the Regge pole parlance.
The constants $H$, $P$ and
 $\alpha$, are  free parameters and  are
determined from the fits.
 We fix $s_0=16.00 ~{\mathrm GeV}^2$, taking a hint from the PDG fit.
PDG adopted the following strategy to fit $\sigma_t$ data.
For the fit to $\sigma^{pp}_t$ the chosen energy range was
from 5 ${\mathrm GeV}$ to cosmic ray regime.
The Froissart-bound-saturating energy dependence is assumed in their
fitting procedure.
Note that in the pre-ISR energy regime the measured cross sections are flat
and measured with very good precision.
Moreover,  Regge pole
contributions, with subleading power behaviors in energy,
 should be included in the pre-ISR energy domain.
However, in the energy
range starting from ISR, the Regge contributions are negligible.
It is worth while to discuss and justify our reasonings for not including the
contributions of subleading
Regge poles to $\sigma_t$ in the energy range starting from ISR point and
beyond (where our interests lie).
Moreover, the subleading Regge pole contributions are  important
in the relatively moderate energy range whiling fitting
 $\sigma_t$  data which remains flat.
 We refer to \cite{eden} and to the review article
of Leader \cite{leader} for detailed discussions.  Let us consider the case
of $pp$ scattering to get a concrete idea.
The Pomeranchuk trajectory contributes a
constant term to $\sigma_t$ and its intercept is $\alpha_P(0)=1$.
Then there are subleading trajectories corresponding to $\omega$, $\rho$,
$A_2$, $\phi$, etc whose energy dependence to the total cross section
is like
 ${ ({ s\over{s_*}})^{\alpha_R(0)-1}}$;  $\alpha_R(0)$ is the intercept
of subleading trajectories such as $\omega$, $\rho$,
$A_2$, $\phi$,
and it is of the order of $1\over 2$.
When   a fit to $\sigma^{pp}_t$ was considered by
Rarita et al \cite{roger}, they concluded, from numerical fits,
that the $\omega$ trajectory dominates \cite{leader,roger} and
the contributions
of other Regge trajectories is quite small \cite{rp}.
They found that the Regge residue (interpreted as the Regge trajectory coupling)
is $R_{pp\omega}\approx 15.5$ mb and $\alpha_{\omega}(0)\approx 0.45$ and
the Regge scale, to define a dimensionless ratio (say
$s\over{s_*}$) is $s_*=1~GeV^2$. Let us estimate what is the
contribution of the $\omega$-trajectory to $\sigma^{pp}_t$ at the ISR energy.
The contribution of the $\omega$-trajectory to $\sigma_t$ is
quite small  in the
 energy range from ISR to LHC. For example, at ISR energy
of $\sqrt{s}=23.5~{\mathrm GeV}$,
the $\omega$-Regge pole contribution to $\sigma_t$ is
approximately  $0.5$ mb  whereas at LHC,
for $\sqrt{s}=8~{\mathrm TeV}$, it is $\approx 0.001$ mb;
the corresponding $\sigma_t$ are $\approx 39~mb$ and $\approx 103~mb$
at $23.5~{\mathrm GeV}$ and $8~{\mathrm TeV}$ respectively. The
parameterizations of \cite{roger} is used for the above estimates.
Consequently, for our purpose, the parameterization
(\ref{ourfit}) is well justified.
We considered the combined data of  $\sigma^{pp}_t$ and $\sigma^{p\bar p}_t$
for
the energy range as mentioned earlier.
The measured values of cross sections against $\sqrt{s}$,
along with the fitted curve, are shown in Fig.~\ref{fig:sigma_vs_sqrt_s}.
The fitted values for the parameters are $P=36.4\pm0.3~mb$, $H=0.22\pm0.02~mb$,
and $\alpha=2.07\pm0.04$.
The quality of the fit, as reflected by the $\chi^2/n.d.f.$ is found
to be moderate due to inclusion of both $\sigma^{pp}_t$ and $\sigma^{p\bar p}_t$ 
measurements from ISR. A fit excluding $\sigma^{p\bar p}_t$ from ISR,
as shown in Fig.~\ref{fig:sigma_vs_sqrt_s} (lower), improves the fit quality
 without
significantly changing the value of the fit parameters.
We have not provided all the references of the experimental papers
from where the data were taken for the plot. The interested reader
may refer to our paper \cite{nm} where detail references are cited.
We find no conclusive evidence for the violation of the  Froissart bound.\\      \begin{figure}[!htbp]
  \centering
  \includegraphics[width=8.6cm]{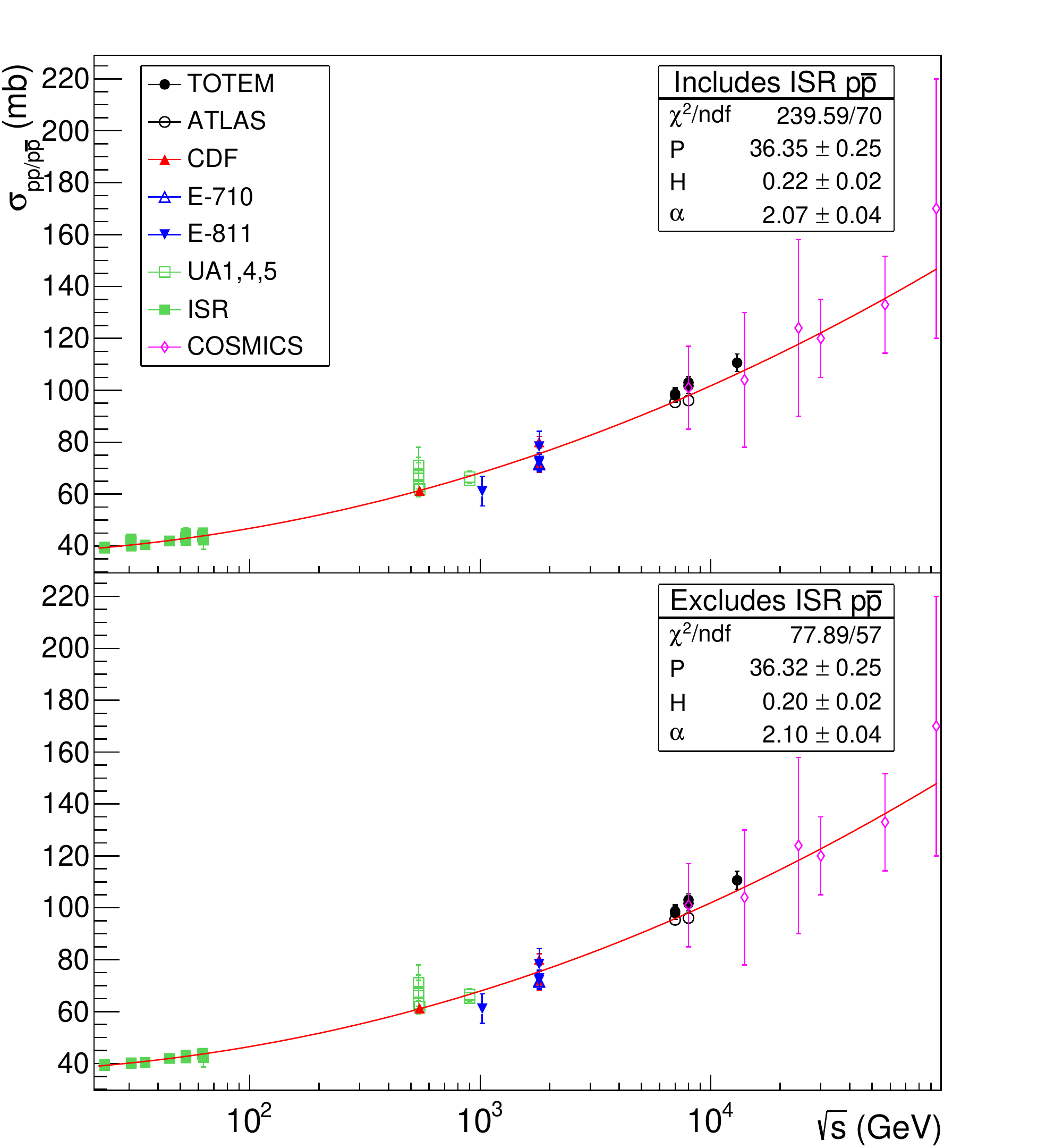}
  \caption{ $\sigma_t^{pp}$ and $\sigma_t^{p\bar{p}}$ against $\sqrt{s}$,
    measured by various experiments.
    The data points are fitted to a function defined in~(\ref{ourfit}).
    The upper plot includes both $pp$ and $p{\bar p}$ data points,
    while the lower one excludes $p{\bar p}$ data points from ISR experiments.
  }
  \label{fig:sigma_vs_sqrt_s}
\end{figure}
We arrive at the conclusion that the data are consistent with
Froissart-Martin bound.
Therefore, there is no indirect evidence for decompactification of
extra dimensions as far as our
proposal goes. If the total cross sections are measured with more
precision showing that there is violation of
Froissart-Martin bound then one might interpret it as a sign of
decompactification. Moreover, our analysis
is in qualitative agreement with the experimental lower bound on
the radius of compactification in the
sense that ATLAS and CMS have not been able to determine the radius
of compactification \cite{tev1,tev2}. We might close this section
with an optimistic note that there is possibility of gathering experimental
evidence in favor of large-radius-compactification
scenario if the precision of the measurements of $\sigma_t$ is improved
significantly. Furthermore, the future high energy accelerators might provide
evidence for the existence of extra spatial dimensions besides discovering
new phenomena hinting at the existence of  fundamental physics.

\bigskip

\noindent{\bf 7. Summary and Discussions}

\bigskip

\noindent We summarize our results in this section and discuss their 
implications.  The objective of the present work is to investigate
the analyticity property of the scattering amplitude in a field theory 
with a compactified spatial dimension on a circle i.e. the $S^1$
compactification. We were motivated to undertake this work from work of
 Khuri \cite{khuri1} who considered
potential scattering with a compact spatial coordinate. He showed the
 lack of analyticity of the forward scattering amplitude under certain 
circumstances. Naturally, it is important to examine what is the
 situation in relativistic field theories. As has been emphasized by
 us before,
lack of analyticity of scattering amplitude in a QFT will be a matter 
of concern since analyticity is derived under very general
axioms in QFT. Thus a compactified spatial coordinate in a theory with 
flat Minkowski spacetime coordinates should not lead
to unexpected drastic violations of fundamental principles of QFT. 
In this paper, initially, a five dimensional neutral massive scalar theory
 of mass, $m_0$, 
was considered in a flat 
Minkowski spacetime.  Subsequently, we compactified a spatial coordinate
 on $S^1$ leading to a spacetime manifold $R^{3,1}\otimes S^1$.
The particles of the resulting theory are a scalar of mass $m_0$ and the 
Kaluza-Klein towers. In this work,
we have focused on elastic scattering of states carrying nonzero 
equal KK charges, $n\ne 0$ to prove fixed-$t$ 
dispersion relations. We have left out the elastic scattering of $n=0$
 states as well as elastic scattering of an $n=0$ state
with an $n\ne 0$ state for nonforward directions. These two cases can 
be dealt with without much problem from our present
work. Moreover, our principal task is to prove analyticity for scattering 
of $n\ne 0$ states and thus complete
the project we started with in order to settle the issue related to 
analyticity as was raised by Khuri  \cite{khuri1} in the context of 
potential scattering. 
We showed in I that forward amplitude satisfies
dispersion relations. However, it is not enough to prove only the 
dispersion relations for the forward  amplitude but a fixed-$t$
dispersion relation is  desirable. We have adopted the LSZ axiomatic 
formulation, as was the case in I, for this purpose. Our results, consequently,
do not rely on perturbation theory whereas, Khuri \cite{khuri1} 
arrived at his conclusions in the perturbative Greens function techniques
as  suitable for a nonrelativistic potential model. Thus the work 
presented here, in some sense, has explored more than 
what Khuri had investigated  in the potential scattering. \\
We have gone through several steps, as mentioned in the discussion 
section of I, in order to accomplish our goal. 
The principal results of this are as follows. First we obtain 
a spectral
representation for the Fourier transform of the causal commutator, $F_c(q)$. 
We discussed the coincidence region which is important
for what followed. In order to identify the singularity free domain, 
we derived analog of the Jost-Lehmann-Dyson theorem. A departure from
the known theorem is that there are several massive states, appearing 
in the spectral representation, and their presence has to be
taken into considerations. Thus, we identified the the singularity free 
region i.e. the boundary of the domain of analyticity. Next, we derived
the existence of the Lehmann ellipse. We  were able to write down 
fixed-$t$ dispersion relations for $|t|$ within the Lehmann ellipse.\\
We have proceeded further. It is not enough to obtain the Lehmann ellipse since the semimajor axis of the ellipse shrinks as $s$ increases.
Thus it is desirable to derive the analog of Martin's theorem \cite{martin1}.
 We appealed to unitarity constraints following Martin and utilized his
arguments on the attributes of the absorptive amplitude and showed that indeed 
Martin's theorem can be proved for the case at hand.
As a consequence, the analog of Froissart-Martin upper bound on total 
cross sections, for the present case, is obtained. The convergence
of partial wave expansions within the Lehmann-Martin ellipse and 
polynomial boundedness for the amplitude, $F(s,t)$ for $|t|$ within 
Lehmann-Martin
ellipse, lead to the Jin-Martin upper bound \cite{jm} for the problem 
we have addressed here. In other words, the amplitude, $F(s,t)$,  does not 
need more than two 
subtractions   to write fixed $t$ dispersion relations for in the domain 
$D_s\otimes D_t$.  \\
We have made two assumptions: (i)  existence of stable particles in 
the entire spectrum of the
the theory defined on $R^{3,1}\otimes S^1$  geometry. Our arguments 
is based on the conservation of KK discrete charge $q_n={{n}\over R}$; 
it is the momentum along the compatified direction.(ii)  The 
absence of bound states.  We have presented some detailed arguments 
in support of (ii).  To put it very concisely, we conveyed that this
flat space  $D=4$
theory with an extra compact $S^1$ geometry results from toroidal 
compactification of five dimensional defined in flat Minkowski space.
In absence of gravity in $D=5$, the lower dimensional theory would not 
have massless gauge field and consequently, BPS type 
states are absent. It is unlikely that the massive scalars 
(even with KK charge) would provide bound states. This is our 
judicious conjecture. \\
We have proposed a novel idea to look for indirect evidence of 
decompactification in the LHC energy regime. As has been elaborated
in Section 6, we argued that precision measurement of very high 
energy total cross section might provide a clue. Suppose, at LHC energy,
the energy dependence of $\sigma_t(s)$ shows a departure from the 
Froissart-Martin bound that total cross section is bounded by $log^2s$.
On the face of it, one might tend to conclude that some of the
axioms of local field theories might not hold. However, on the
other hand, if an extra spatial dimension decompactifies then
the generalized Froissart-Martin bound in $log^{D-2}s$ where $D$ is
the number of spacetime dimensions \cite{jmjmp}. Therefore,
in the event of such an observation, we need not question the
fundamental axioms. We have fitted the data \cite{nm} from ISR energies to the
LHC energy and included the cosmic ray data points for $\sigma_t(s)$.
We kept the power of $log s$ as a floating parameter. Our analysis
does not indicate conclusive violation of the Froissart-Martin bound
\\
Another interesting aspect needs further careful consideration. 
Let us start with a five dimensional Einstein theory minimally coupled to 
a massive
neutral scalar field of mass $m_0$. We are unable to fulfill requirements 
of LSZ axioms in the case of the five dimensional theory in curved spacetime. 
Furthermore,
let us compactify this theory to a geometry $R^{3,1}\otimes S^1$.
 Thus the resulting scalar field lives in flat Minkowski space with a
 compact
dimension. We have an Abelian gauge field in $D=4$, which arises 
from $S^1$ compactification of the 5-dimensional Einstein metric. The 
spectrum
of the theory can be identified: (i) There is a massive scalar of mass 
$m_0$ descending of $D=5$ theory accompanied by KK tower of states.
(ii) A massless gauge boson and its massive KK partners. 
(iii) If we expand the five dimensional metric around four dimensional 
Minkowski
metric when we compactify on $S^1$, we are likely to have massive 
spin 2 states (analog of KK towers). We may construct a Hilbert space
in $D=4$ i.e with geometry $R^{3,1}\otimes S^1$.  It will be interesting 
to investigate the analyticity properties of the scattering amplitudes and
examine the high energy behaviors. Since only a massless spin 1 particle 
with Abelian gauge symmetry appears in the spectrum, it looks as
if the analyticity of amplitudes will not be affected. However, there might be
 surprises since a massive spin 2 particle is present in the spectrum. 
    
\bigskip

\noindent { \bf Acknowledgments}: I dedicate this article to the fond
memories of Andr\'e Martin. The monumental contributions of Andr\'e to
establish  analyticity properties of scattering amplitudes have been 
a source of inspirations. The results presented in this article are
an outcome of prolonged discussions with him. I have benefited 
considerably from discussions with Stefan Theisen over last few years.
I thank Hermann Nicolai for several discussions and critical comments
on my works on study of analyticity.
It has been a great pleasure to collaborate with Aruna Kumar Nanyak. I am 
thankful to Ignatious Antoniadis and Luis Alvarez Gaume for valuable
discussions. I thank Hermann Nicolai and Max-Planck-Institute for Gravitational
Physics (Albert Eistein Institute), Golm, for their very gracious and warm 
hospitalities during several visits (2015-2019). I thank Dieter L\"ust and
Max-Planck-Institute for Physics ( Werner Heisenberg Institute), Munich, for 
their warm hospitality where a part of this work was done.

\end{document}